\documentclass[global]{svjour}
\usepackage{graphicx}
\usepackage{epsfig}
\usepackage[english]{babel}
\usepackage[utf8x]{inputenc}
\usepackage{amsmath}
\usepackage{amsfonts}
\usepackage{dsfont}
\usepackage{longtable}
\usepackage{array}
\newcolumntype{L}[1]{>{\raggedright\let\newline\\\arraybackslash\hspace{0pt}}m{#1}}

\begin{document}

\title{Comparison of Gaussian and super Gaussian laser beams for addressing atomic qubits}

\author{Katharina Gillen-Christandl\inst{1} \and Glen D. Gillen\inst{1} \and M. J. Piotrowicz\inst{2,3} \and M. Saffman\inst{2}}

\institute{Physics Department, California Polytechnic State University, 1 Grand Avenue, San Luis Obispo, CA 93407, USA \and Department of Physics, University of Wisconsin-Madison, 1150 University Avenue, Madison, Wisconsin 53706, USA \and Department of Physics, University of Michigan, Ann Arbor, MI 48109, USA}

\date{\today}

\begin{abstract}
We study the fidelity of single qubit quantum gates performed with two-frequency laser fields that have a Gaussian or super Gaussian spatial mode. Numerical simulations are used to account for imperfections arising from atomic motion in an optical trap,  spatially varying Stark shifts of the trapping and control beams, and transverse and axial misalignment of the control beams. Numerical results that account for the three dimensional distribution of control light show that a super Gaussian mode with intensity $I\sim e^{-2(r/w_0)^n}$ provides reduced sensitivity to atomic motion and beam misalignment. 
Choosing a super Gaussian with $n=6$  the decay time of finite temperature Rabi oscillations can be increased by a factor of 60 compared to an $n=2$ Gaussian beam, while reducing crosstalk to neighboring qubit sites.  
\end{abstract}

%\pacs{}
%\keywords{}

\maketitle
%\tableofcontents

\section{Introduction}

Atomic qubits encoded in hyperfine ground states are one of several approaches being developed for quantum computing experiments\cite{Ladd2010}. Single qubit rotations can be performed with microwave radiation or two-frequency laser light driving stimulated Raman transitions. The microwave approach, while simpler in implementation, does not directly yield site resolved operations in a multi-qubit array. Single site selectivity can be achieved with microwaves using a magnetic field gradient\cite{Schrader2004},  or tightly 
focused Stark shifting beams\cite{Xia2015,YWang2015,Lee2013,Labuhn2014},  or with focused two-frequency Raman light\cite{Yavuz2006,Knoernschild2010}.

In order to achieve as high a fidelity as possible for qubit rotations the field strength must be precisely controlled at the location of the atom. 
Finite temperature position fluctuations or jitter in the alignment of optical beams lead to  gate errors due to variation of the optical intensity interacting with the atom. There is also a second cause of gate errors due to variations in the differential Stark shift experienced by an atom as a function of location in an optical trap.
Both the spatially varying trap intensity, and the spatially varying Raman intensity lead to a variable Stark shift, and hence position dependent qubit detuning errors. While these errors can be minimized by cooling to the motional ground state of the trap, ground state cooling tends to take longer than cooling to a thermal state, and any heating sources lead to motional excitation. 

In light of these effects it is of interest to design the experimental control system to minimize errors due to atomic motion or beam misalignment.
It is known that using beams with more uniform intensity profiles can lead to improved fidelity of Rabi oscillations\cite{Reetz-Lamour2008b,Huber2011} and that specially shaped beams have the potential for reduced crosstalk\cite{Saffman2004}. 
 In this paper we present a detailed analysis of the fidelity of qubit rotations driven by two-frequency Raman light. We investigate how modifying the beam shape from the typical Gaussian profile of a laser beam to a super Gaussian with a ``top-hat" like profile can reduce the effects of intensity variations and simultaneously minimize crosstalk to neighboring qubit sites. 
Numerical simulations are used to quantify the effects of the thermal motion of  trapped atoms as well as radial and axial control beam misalignment on the fidelity of Rabi oscillations with finite temperature atoms. 
We show that the super Gaussian beam  can dramatically reduce both motional dephasing and crosstalk to neighboring sites. While we specifically analyze the case of two-frequency Raman light we expect that our results will also be applicable to localized gates that rely on microwaves with focused Stark shifting beams. 

In Sec.~\ref{sec:theory}, we describe how to calculate the three-dimensional (3D) intensity profiles of Gaussian and super Gaussian beams, define the density weighted intensity variance, discuss the optical trap geometry used in our simulations and review how the two-photon Rabi transitions in $^{133}$Cs depend on the position of the atoms due to the intensity profiles of the trapping and control lasers. In Sec.~\ref{sec:calcs}, we present our calculation results for the 3D intensity profile of super Gaussian beams as well as the effects of atom temperature and control beam misalignment on the density weighted intensity variance and the Rabi oscillations of the atomic qubits, before summarizing our conclusions in Sec.~\ref{sec:conclusions}.

\section{Theory}
\label{sec:theory}

In this section, we theoretically investigate the effects of using a super Gaussian beam to address the atoms instead of a TEM$_{00}$ Gaussian beam.  First, we define what constitutes a super Gaussian beam for this investigation. Then we present a model to calculate the propagation of the super Gaussian beam in order to have a complete three-dimensional (3D) mapping of the electric field and intensity distributions. In Sec.~\ref{sec:calcs}, we use the 3D intensity distributions to quantitatively compare the spatial intensity variations between Gaussian and super Gaussian beams and their subsequent effects on Rabi oscillations of trapped atoms to compare the evolution of the qubit states for super Gaussian addressing beams to TEM$_{00}$ addressing beams.  

The position-dependent variations in the differential AC Stark shifts result in reduced flopping amplitudes, faster decay of the Rabi oscillations, and changes to the Rabi frequency due to position-dependent variations in the electric field of the light experienced by the atom. The dominant contribution to the differential AC Stark shift at the location of each trap site is that of the addressing laser beam. Thus, to eliminate the position-dependent differential AC Stark shifts of the atomic states, one would ideally use an addressing beam with a uniform intensity over the entire trap volume; i.e., a flat-top beam profile.  However, for qubit operations tight focusing is required to address a single qubit without crosstalk at neighboring  qubit sites.  Also, the sharp spatial features required for flat-top beams in the focal plane result in undesired spatial oscillations away from the focal plane. It is therefore necessary to consider both the uniformity of the intensity and spatial crosstalk when selecting an optimized beam profile. 

\subsection{Gaussian vs. super Gaussian beams}

\label{sec:SGbeams}

 If the addressing beam is a TEM$_{00}$ Gaussian beam then the electric field of the beam can be analytically modeled in three dimensions using \cite{Gillen2013}
\begin{eqnarray}
E \left( x, y, z, t \right) & = &  E_0 \frac{w_0}{w \left( z \right)}
\exp\left[ {- {\displaystyle  \frac{\left( x^2 + y^2 \right)}{w^2 \left( z \right)}} }  \right] \nonumber \\  
&& \times \exp \left( {i {\displaystyle  \left[ \frac{k}{2} \frac{\left( x^2 + y^2 \right)}{R \left( z \right)} -\tan^{-1}  \frac{z}{z_0} \right]} } \right) e^{i\left(kz-\omega t\right)} ,
\label{eq:TEM00scalarE}
\end{eqnarray}
where $E_0$ is the amplitude of the electric field at the center of the focal plane, or $x=y=z=0$. Because $I \propto |E|^2$, the corresponding three-dimensional intensity distribution is 
\begin{equation}
I \left( x, y, z \right) = \frac{I_0} {1 + (z/z_0)^2} \exp \left[-2 \frac{\left( x^2 + y^2 \right)}{w^2 \left( z \right)} \right] ,
\label{eq:Igaussxyz}
\end{equation}
where $I_0$ is the intensity at the center of the focal plane, the Gaussian width (radius where the intensity is $e^{-2}$ of the peak value) as a function of axial distance from the plane of the beam waist, $w \left( z \right)$, is $w \left( z \right) = w_0 \sqrt{\displaystyle { \left( 1 + \frac{z^2}{z_0^2} \right)}} ,$
where $z_0=\pi w_0^2 / \lambda$ is the Rayleigh Range,  and $w_0$ is the Gaussian beam waist in the focal plane, and the beam radius of curvature is $R(z)=z(1+z_0^2/z^2)$.

Super Gaussian beams are light patterns whose intensity profiles reside in the regime between smoothly propagating TEM$_{00}$ Gaussian beams and pure flat-top beams.  A super Gaussian beam is defined here as one whose intensity profile at the beam waist follows the mathematical function \cite{Lu1996}
\begin{equation}
I \left( r \right) = I_0 e^{-2 \left( \frac{r}{w_0} \right)^n} \mbox{, where } n > 2 ,
\label{eq:SGgeneral}
\end{equation}
and $n$ is the order of the super Gaussian.  We also define the phase front in the $z=0$ plane to be planar and perpendicular to the propagation direction.  For super Gaussian beams the axial location of the peak intensity, or the axial location of the narrowest beam distribution, is not the same as the location of the planar wave front which we refer to as the ``beam waist'', the focal plane, or the $z=0$ plane; as is discussed further in Sec.~\ref{sec:SGcalcs}.

\begin{figure}[!t]
 \centering
 \includegraphics[width=70mm]{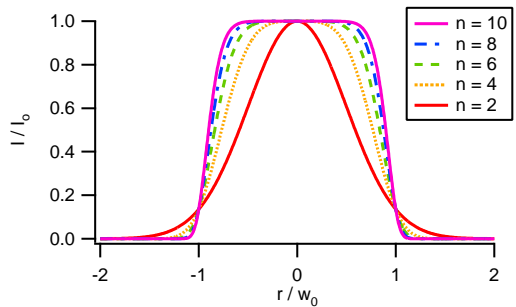}
\caption{(Color online) Radial intensity profiles for super Gaussian beams of orders $n=$2, 4, 6, 8, and 10.}
\label{fig:SGvsn}
\end{figure}

If the super Gaussian has order 2 then the beam profile is that of a TEM$_{00}$ Gaussian beam in the focal plane, and the beam propagates exactly as a TEM$_{00}$ Gaussian beam would for axial locations away from the beam waist plane.  As the order $n$ of the super Gaussian increases, the effect on the beam waist profile is to widen and flatten the central intensity peak of the beam while increasing the rate of change of the intensity of the sides of the beam as illustrated in Fig.~\ref{fig:SGvsn}. As we show in the following a super Gaussian with $n=6$ provides significant improvement of gate fidelity and reduction of crosstalk. Although the intensity difference
between the $n=6$ beam and $n=4$ or $n=8$ is not more than $\sim 20\%$ the performance improves by a much larger factor. It is therefore important to accurately prepare the desired beam profile. We discuss possible methods for doing so in Sec. \ref{sec:conclusions}. 

Unlike TEM$_{00}$ Gaussian beams where the electric field at any point in space can be analytically determined using Eq. (\ref{eq:TEM00scalarE}), an analytical form for super Gaussian beams in all space does not  exist.  The analytical form given in Eq. (\ref{eq:SGgeneral}) is only valid for points within the plane containing the beam waist, or $\left( x, y, 0 \right)$.  For any points outside of the beam waist plane the electric field at the point of interest must be determined using beam propagation methods and numerical integration.  Scalar diffraction theory can be used to calculate the propagated electric field behind the focal plane.  The model chosen here for determining the electric field at the point of interest is the Rayleigh-Sommerfeld diffraction integral \cite{Gillen2013,Gillen2004}.  Rayleigh-Sommerfeld diffraction is a scalar diffraction model which integrates the known field values over an input plane, $\left( x_0, y_0, z_0 = 0 \right)$, and propagates the field to a particular point of interest, $\left( x, y, z\right)$ using
\begin{equation}
E \left( x, y, z \right) = \frac{k z}{i 2 \pi} \int \!\!\! \int E_{z_0 = 0}
\frac{e^{ik \rho}}{\rho^2} \left( 1 - \frac{1}{i k \rho}
\right) \,dx_0 \,dy_0 ,
\label{eq:RSdiffintegral}
\end{equation}
where $k$ is the wave number, and $\rho$ is the distance from the integration point to the point of interest, or
\begin{equation}
\rho \left(x_0, y_0, z_0 = 0, x, y, z \right) = \sqrt{(x - x_0)^2 + (y - y_0)^2 + z^2} .
\nonumber%\label{eq:Rho}
\end{equation}

In this investigation, it is assumed that the input plane for the Rayleigh-Sommerfeld diffraction integral coincides with the axial location of the beam waist, and that the phase front of the electric field is planar with the direction of energy propagation to be the $+z$-direction for all points within the beam waist plane.  Using these assumptions, the known electric field term in Eq. (\ref{eq:RSdiffintegral}) can be simply written as
\begin{equation}
E_{z_0 = 0} = E_0 e^{- \left( \frac{r}{w_0} \right)^n} .
\label{eq:SGE0}
\end{equation}
Substitution of Eq. (\ref{eq:SGE0}) into Eq. (\ref{eq:RSdiffintegral}) and numerically integrating over the input plane for a desired point of interest with $z > 0$ one obtains the scalar value of the electric field  at the point of interest.

\subsection{Density weighted intensity variance}
\label{sec:intvar}
If the intensity of the addressing beam is not uniform over the entire volume of the trap site then as the atom moves within the trapping volume it samples different intensity values resulting in position-dependent variations in the differential AC Stark shifts between atomic states and Rabi frequency of atomic transitions. In order to help quantify the total variation of intensities an atom at a particular temperature would be exposed to we can calculate a density weighted intensity variance (which we will subsequently refer to as the ``intensity variance''), or $\sigma_{I \Psi}$. We quantitatively define the intensity variance to be
\begin{equation}
\sigma_{I \Psi}^2 = \int{ I_n^2 \left| \Psi \right|^2 dv} - \left[ \int{ I_n \left| \Psi \right|^2 dv}  \right]^2 ,
\label{eq:ABIVgeneral}
\end{equation} 
where the integration is performed over all space and $I_n\left( x ,y, z \right)= \frac{I \left( x ,y, z \right)}{I_0}$ is the normalized position-dependent intensity of the addressing beam, $I_0$ is the peak intensity in the input beam waist plane, and $\Psi$ is the normalized wavefunction of the harmonically trapped atom with temperature-dependent widths $\sigma_x$, $\sigma_y$, and $\sigma_z$, as discussed in the next section.  By inspection, we see that if the intensity distribution, $I$, is a uniform constant for all space where the wavefunction $\Psi$ is non-zero then $\sigma_{I \Psi}$ would be zero.  Thus, the higher the variation of intensity the atom samples during the time which it is exposed to the addressing beam, the larger the value of $\sigma_{I \Psi}$.

\subsection{Dipole trap formed by Gaussian beams}
\label{sec:theorysetup}

For the detailed analysis of the intensity variance, crosstalk, and Rabi oscillations we have assumed blue-detuned 
optical traps using the 49 site Gaussian beam array introduced in \cite{Piotrowicz2013}. Each trapping site  is formed by the intensity overlap of four Gaussian beams, each of power $P$,  in a unit cell of the array with area $d\times d$. The trapping potential is 
\begin{equation}
U_T({\bf r})=-\frac{\alpha}{2\epsilon_0 c}I_T
\label{eq:trappotential}
\end{equation}
 where $\alpha$ is the atomic polarizability, and $I_T$ is the intensity found by adding the contributions from the four beams forming a unit cell of the array. 
Following the  analysis in \cite{Piotrowicz2013,SZhang2011} we find the trap spring constants
\begin{subequations}
\begin{align}
\kappa_x & = \frac{32\left|U_d\right|}{\pi d^2}s^4 \left( s^2 - 1 \right)e^{-s^2} , \\
\kappa_y & = \kappa_x, \\
\kappa_z & = \frac{16\lambda^2\left|U_d\right|}{\pi^3 d^4}s^6 \left( s^2 - 1 \right)e^{-s^2} . 
\end{align}
\label{eq:kappaxyz}
\end{subequations}
Here $s=d/w_0$ with $d$ the array period, $w_0$ the waist of each trapping beam,  $U_d=\frac{\alpha}{2\epsilon_0 c}I_d$, and $I_d=P/d^2$ is the average intensity of each unit cell. 
The corresponding trap frequencies are $\omega_j=\sqrt{\frac{\kappa_j}{m_a}}$, where $m_a$ is the atomic mass, and $j=x,y,z$.
Using the relationship $\frac{1}{2}\kappa_j\sigma_j^2=\frac{1}{2} k_B T$, the time-averaged position variances are
\begin{subequations}
\begin{align}
\sigma_x^2 & = \sigma_{x0}^2 \frac{e^{s^2}}{s^4 \left( s^2 - 1 \right)} , \\
\sigma_y^2 & = \sigma_x^2, \\
\sigma_z^2 & = \sigma_{z0}^2 \frac{2 e^{s^2}}{s^6\left(s^2-1\right)} . 
\end{align}
\label{eq:sigmaxyz}
\end{subequations}
Equations (\ref{eq:kappaxyz}) and (\ref{eq:sigmaxyz}) are the same as Eqs. (10) and (11) from \cite{Piotrowicz2013}, except for corrections to Eqs. (10c) and (11c).
The parameters $\sigma_{x0}$ and $\sigma_{z0}$ are defined as
\begin{equation}
\sigma_{x0} = \sqrt{\frac{\pi d^2 k_B T}{32 \left| U_d \right|}}, ~~~
\sigma_{z0} = \sqrt{\frac{\pi^3 d^4 k_B T}{32 \lambda^2 \left| U_d \right|}},
\nonumber%\label{eq:sigma0z}
\end{equation}
where $k_B$ is Boltzmann's constant and $T$ is the atom temperature.
For our calculations, we used a trap spacing of $d=3.8~\mu$m, a $\lambda=780~\rm nm$ trap wavelength,  a  laser beam waist of 1.73~$\mu$m (making the normalized array period $s=2.197$), and a laser power of 3~W, split into 64 equal laser beams so $P=0.047~\rm W$. The polarizability $\alpha_{780}$ for the 780-nm trap light was calculated using a standard sum over states expression\cite{LeKien2013} including the 
$6P_{1/2,3/2}$ and $7P_{1/2,3/2}$ levels in Cs. Using numerical values from \cite{Sansonetti2009,Iskrenova-Tchoukova2007} for the transition wavelengths and dipole matrix elements we find $\alpha_{780, \rm cgs}=-250.\times 10^{-24}~\rm cm^3$ in cgs units and in SI units $\alpha_{780,\rm SI}=(4\pi\epsilon_0) \alpha_{780,\rm cgs}$.

\subsection{Rabi oscillations}
\label{sec:Rabioscillations}

A two-level atom interacting with a monochromatic field undergoes Rabi oscillations between its two levels. We will designate the two states as ground (lower) state $|g\rangle$ and excited (upper) state $|e\rangle$ with energies $\hbar\omega_g$ and $\hbar\omega_e$, respectively. 
Writing the state vector as  $|\psi\rangle=c_g(t)e^{-\imath \omega_g t} |g\rangle + c_e(t)e^{-\imath \omega_e t} |e\rangle $ 
 the Schr\"odinger equation in the rotating wave approximation takes the form  
\begin{subequations}
\begin{eqnarray}
\frac{d c_g}{dt}   &=&  i \frac{\Omega^*}{2} c_e  e^{\imath \Delta t },\\
\frac{d c_e}{dt}  &=& i \frac{\Omega}{2} c_g  e^{-\imath \Delta t },
\end{eqnarray}
\label{eq:coupledRabi}
\end{subequations}
with $\Delta=\omega-\omega_{eg},$ $\omega$ is the optical frequency, $\omega_{eg}=\omega_e-\omega_g$, the  Rabi frequency is $\Omega = d_{eg}{\mathcal E}/\hbar$ where  $d_{eg}=-e\langle e|\hat{r}|g\rangle$ is the  matrix element of the dipole operator $\hat{d}=-e\hat{r}$ , and $e$ is the elementary charge.

Solving this system of coupled differential equations results in the solution 
\begin{equation}
\left(\begin{array}{c}c_g(t) \\c_e(t)\end{array}\right)=\mathbf{M}\cdot\left(\begin{array}{c}c_g(t_0) \\c_e(t_0)\end{array}\right),
\label{eq:statevector}
\end{equation}
where $t_0$ is the initial time and
{\small 
\begin{eqnarray}
\mathbf{M}&=&
\left(
\begin{matrix}
e^{\imath \frac{\Delta (t-t_0)}{2}}\left[ \cos\left( \frac{\Omega' (t-t_0)}{2} \right) -i \frac{\Delta}{\Omega'}\sin\left( \frac{\Omega' (t-t_0)}{2} \right) \right] &
 i e^{\imath \frac{\Delta (t+t_0)}{2}}\, \frac{\Omega^*}{\Omega'}\sin\left( \frac{\Omega' (t-t_0)}{2} \right) 
\cr 
i e^{-\imath \frac{\Delta (t+t_0)}{2}}\,  \frac{\Omega}{\Omega'}\sin\left( \frac{\Omega' (t-t_0)}{2} \right)&
e^{-\imath \frac{\Delta (t-t_0)}{2}}\left[ \cos\left( \frac{\Omega' (t-t_0)}{2} \right) +i \frac{\Delta}{\Omega'}\sin\left( \frac{\Omega' (t-t_0)}{2} \right) \right]
\end{matrix}
\right)\nonumber\\
\label{eq:rotationmatrix} 
\end{eqnarray}
}
with the effective off-resonance Rabi frequency $\Omega'=\sqrt{|\Omega|^2 + \Delta^2}$.
When the atom is initially in the ground state, the time dependent probabilities to be in the ground and excited states are found to be
\begin{eqnarray}
|c_g(t)|^2 
&=&   \cos^2\left( \frac{\Omega' t}{2} \right)+\frac{\Delta^2}{|\Omega|^2+\Delta^2}\sin^2\left( \frac{\Omega' t}{2} \right),\nonumber\\
|c_e(t)|^2 
&=&   \frac{|\Omega|^2}{|\Omega|^2+\Delta^2}\sin^2\left( \frac{\Omega' t}{2} \right).
\label{eq:populations}
\end{eqnarray}

We see that the ground and excited state probabilities undergo Rabi oscillations with the effective off-resonance Rabi frequency $\Omega'$.

\subsection{Rabi oscillations due to Raman transitions in $^{133}$Cs}

In this work, we investigate the Rabi oscillations due to a two-photon stimulated Raman transition from the $6^2S_{1/2},F=3, m_F=0$ state of $^{133}$Cs, corresponding to the lower state $|g\rangle$ from Sec.~\ref{sec:Rabioscillations}, to $6^2S_{1/2}, F=4, m_F=0$, corresponding to the upper state $|e\rangle$ from Sec.~\ref{sec:Rabioscillations} via the $7^2P_{1/2}$ state driven by a pair of 459 nm Raman laser beams. In order to calculate the evolution of the $F=4$ population with time, we need to use the appropriate on-resonance Rabi frequency $\Omega$ and detuning $\Delta$ for this two-photon transition. As shown in the Appendix, the two-photon Rabi frequency for a $\Lambda$ type coupling scheme with resolved excited state hyperfine structure is
\begin{equation}
\Omega=\frac{\Omega_{1,0}\Omega_{2,0}^*}{32}\left(\frac{1}{ \Delta_R-\Delta_{F'3}}+\frac{5/3}{\Delta_R-\Delta_{F'4}}\right),
\label{eq:Rabifrequency}
\end{equation}
with 
\begin{equation}
\Omega_{i,0}= {\mathcal E}_i e \langle 7^2P_{1/2}||r||6^2S_{1/2}\rangle/\hbar,
\nonumber%\label{eq:Omega0}
\end{equation}
where $i=1,2$ refers to Raman laser beams 1 and 2, respectively,
$\Delta_R$ is the detuning of the first Raman laser from the $6^2S_{1/2}, F=3\rightarrow7^2P_{1/2}$ (fine structure level) transition, and $\Delta_{F'3}=-2\pi\times212.3~\text{MHz}$ and $\Delta_{F'4}=2\pi\times165.1~\text{MHz}$ are the hyperfine shifts from the $7^2P_{1/2}$ fine structure level to the $F'=3$ and 4 hyperfine states, respectively. ${\mathcal E}_{1,2}$ refers to the electric field amplitudes of Raman lasers 1 and 2, respectively. $e\langle 7^2P_{1/2}||\hat{r}||6^2S_{1/2}\rangle=0.276 e a_0$ is the reduced dipole matrix element for the $6^2S_{1/2}\rightarrow7^2P_{1/2}$ transition in $^{133}$Cs \cite{Iskrenova-Tchoukova2007}, and $a_0$ is the Bohr radius.

As we discuss in the next section, the detuning, $\Delta$, for our calculation will be due to changes in the differential AC Stark shift between the $6^2S_{1/2}, F=3$ and $F=4$ hyperfine ground states. There are two contributions to the differential Stark shift, one from the Raman addressing beams, $\Delta_{acR}$, and one from the trap light, $\Delta_{acT}$, so
\begin{equation}
\Delta_{ac}=\Delta_{acR}+\Delta_{acT}.
\label{eq:diffStarktotal}
\end{equation}

Since the Raman laser beams are near resonant to the $6^2S_{1/2}\rightarrow7^2P_{1/2}$ transition, the differential AC Stark shift due to the Raman beams is (see the Appendix for the derivation)
\begin{eqnarray}
\Delta_{acR}=\frac{|\Omega_{1,0}|^2}{64} \left(\frac{1}{\Delta_R-\Delta_{F'3}+\Delta_{hf}}\right.&+&\left.
\frac{5/3}{\Delta_R-\Delta_{F'4}+\Delta_{hf}}\right.\nonumber\\
-\left.\frac{1}{\Delta_R-\Delta_{F'3}-\Delta_{hf}}\right.&-&\left.
\frac{5/3}{\Delta_R-\Delta_{F'4}-\Delta_{hf}}\right).
\label{eq:diffStarkRaman}
\end{eqnarray}
where $\Delta_{hf}=2\pi\times9.192631770$~GHz is the ground state hyperfine splitting in $^{133}$Cs. Here, we have assumed that both Raman beams have the same power, waist, and alignment, so that $\Omega_{1,0}=\Omega_{2,0}$.
For the differential Stark shift due to the far-detuned 780-nm trap laser, we use the expression 
\begin{equation}
\Delta_{acT}=-\frac{\alpha}{4}\frac{\Delta_{hf}}{\Delta_{T}}\frac{\left|{\mathcal E}_{T}\right|^2}{\hbar},
\label{eq:diffStarktrapE}
\end{equation}
where $\alpha$ is the polarizability of the $6^2S_{1/2}$ state in $^{133}$Cs in SI units and ${\mathcal E}_{T}$ is the electric field amplitude of the trap light. $\Delta_T$ is the effective detuning of the trap laser from the D1 and D2 transitions in $^{133}$Cs, given 
by \cite{Kuhr2005,Schrader2001,Grimm2000} $\frac{1}{\Delta_{T}}=\frac{1}{3}\frac{1}{\Delta_{D1}}+\frac{2}{3}\frac{1}{\Delta_{D2}}$ with $\Delta_{D1}=2\pi\left(\frac{c}{780\text{nm}}-\frac{c}{894\text{nm}}\right)$ and $\Delta_{D2}=2\pi\left(\frac{c}{780\text{nm}}-\frac{c}{852\text{nm}}\right)$.

\subsection{Position dependence of Rabi oscillations}
\label{sec:positiondependenceRabi}

Thus far, we have assumed that the intensity of the light field and the atomic energy levels are uniform throughout space. However, the electric field amplitudes $\mathcal{E}_1, \mathcal{E}_2,$ and $\mathcal{E}_T$, depend on the position of the atom in the Raman beams and in the trap, so
\begin{equation}
\mathcal{E}_{1,2,T}=\mathcal{E}_{1,2,T}\left(x,y,z\right).
\nonumber%\label{eq:Efieldxyz}
\end{equation}

Therefore, the effective on-resonance Rabi frequency $\Omega$ and the differential AC Stark shifts $\Delta_{acR}$ and $\Delta_{acT}$ due to the Raman lasers and the trap light, respectively, depend on the position of the atom. Consequently, as the atom moves through the trap, $\Omega$ and $\Delta$ will change with time.
Thus, to calculate the Rabi oscillations of an atom in our system, Eq. (\ref{eq:statevector}) must be solved numerically in small time steps to account for these changes with time as the atoms are moving.

The electric field strength $|\mathcal{E}|$ is related to the  light intensity $I\left(x,y,z\right)$ at the location of the atom by 
$\left|\mathcal{E}\left(x,y,z\right)\right|=\sqrt{\frac{2I\left(x,y,z\right)}{\epsilon_0 c}}.$
For Gaussian addressing beams of waist $w_0$, this intensity is calculated from the laser power $P$ at the atoms as
\begin{equation}
I \left( x, y, z \right) =I_0 I_n\left(x,y,z\right)
\label{eq:Ixyznormint}
\end{equation}
with $I_0=\frac{2P}{\pi w_0^2}$ and
\begin{equation}
I_n \left( x, y, z \right) = \frac{1} {1 + (z/z_o)^2} \exp \left[-2 \frac{\left( x^2 + y^2 \right)}{w^2 \left( z \right)} \right].
\nonumber%\label{eq:normintgauss}
\end{equation}

For the super Gaussian addressing beams, we used the same maximum intensity, $I_0$, as for the Gaussian beams, multiplied by the numerically calculated normalized intensity profiles, $I_n\left(x,y,z\right)$, as described in Sec.~\ref{sec:SGbeams}.

In the calculations, we used Raman laser beams of identical power and waists that are perfectly aligned with each other, so $\mathcal{E}_1=\mathcal{E}_2$. From Eqs. (\ref{eq:Rabifrequency},  \ref{eq:Ixyznormint}) we thus find the position-dependent on-resonance Rabi frequency for evaluating Eq. (\ref{eq:statevector}) to be
\begin{equation}
\Omega\left(x,y,z\right)=\Omega\left(0,0,0\right)I_n\left(x,y,z\right).
\nonumber%\label{eq:Rabifrequencyxyz}
\end{equation}
Because the atoms are moving, for each time step the position of the atom must be calculated and the on-resonance Rabi frequency for that atom determined in order to evaluate Eq. (\ref{eq:statevector}).

By the same means, using Eq. (\ref{eq:diffStarkRaman}), we find that the differential Stark shift due to the Raman laser beams is
\begin{equation}
\Delta_{acR}\left(x,y,z\right)=\Delta_{acR}\left(0,0,0\right)I_n\left(x,y,z\right).
\nonumber%\label{eq:diffStarkRamanxyz}
\end{equation}

The position-dependent differential Stark shift due to the trap light can be obtained using $U_T=-\frac{1}{4}\alpha \left|\mathcal{E}_T\right|^2$ and Eq. (\ref{eq:diffStarktrapE}) as
\begin{equation}
\Delta_{acT}\left(x,y,z\right)=\frac{\Delta_{hf}}{\Delta_{T}}\frac{U_T\left(x,y,z\right)}{\hbar},
\nonumber%\label{eq:diffStarktrapU}
\end{equation}
leading to a total differential Stark shift of
\begin{equation}
\Delta_{ac}\left(x,y,z\right)=\Omega\left(0,0,0\right)I_n\left(x,y,z\right)+\frac{\Delta_{hf}}{\Delta_{T}}\frac{U_T\left(x,y,z\right)}{\hbar}.
\nonumber%\label{eq:diffStarktotalxyz}
\end{equation}

As for the detuning $\Delta$ from Eq. (\ref{eq:rotationmatrix}), we tune the Raman lasers exactly to the $6^2S_{1/2},F=3$ to $F=4$ hyperfine ground state transition, taking into account the AC Stark shifts from the trap and Raman lasers for an atom at the center of the trap and assuming the Raman laser beams are perfectly centered. Therefore, $\Delta$ will be the difference of the differential Stark shift of an atom at position $x,y,z$ from that at the center of the trap and addressing laser beams. Since the atoms are moving, this needs to be evaluated at each time step.

\section{Calculations and simulations}
\label{sec:calcs}

\subsection{Super Gaussian beam propagation}
\label{sec:SGcalcs}

In this manuscript we want to duplicate the experimental setup used in Ref. \cite{Piotrowicz2013} and investigate the effects of using super Gaussian beams for the addressing laser instead of a TEM$_{00}$ Gaussian beam.  Therefore, the parameters used for the computations  for the TEM$_{00}$ Gaussian beam largely match those used in the experiment.  For this reason, the width of the Gaussian beam, $w_{n=2}$, for this investigation was chosen to be 2.30~$\mu m$. As observed in Fig.~\ref{fig:SGvsn}, when the super Gaussian order $n$ increases both the region of uniform intensity increases as well as the magnitude of the ramp rate of the intensity outside of the central region.  For radial values of $r < w$ the intensity of the beam increases with the order $n$, and for radial values of $r>w$ the intensity of the beam decreases as the order $n$ increases.  There is a trade-off to be considered when deciding what width to use for each super Gaussian beam order: A larger width increases the size of the central uniform intensity volume, but may increase the crosstalk intensity at the location of a neighboring site in the case of radial misalignment.  

All experimental beams have radial and axial misalignment, or jitter, due to experimental conditions; i.e., mechanical vibrations in the optical equipment, air currents in the lab room, etc.  If the beam widths for $n>2$ super Gaussian beams are chosen such that the crosstalk intensity is equal to that of $n=2$ at $r=d$, then all super Gaussian beams will have much higher variations in the crosstalk intensity due to radial jitter and the steeper intensity ramp rates, an undesired outcome.  Thus, we have set the widths of the super Gaussian beams such that the crosstalk intensity for each value of $n$ is less than, or equal to, that of a centered TEM$_{00}$ Gaussian beam up to a maximum acceptable value of radial jitter, $r_0$; i.e., the worst-case crosstalk scenario for a misaligned super Gaussian beam is equal to the best-case scenario of a centered Gaussian beam, or
\begin{equation}
I_{n>2}\left(r=d-r_0, z=0\right)= I_{n=2}\left(r=d, z=0\right),
\nonumber%\label{eq:In2}
\end{equation}
or, using Eq. (\ref{eq:SGgeneral}),
\begin{equation}
w_n=\left(\frac{w_0}{d}\right)^{2/n} \left(d-r_0\right) ,
\label{eq:wn}
\end{equation}
where $w_0$ is the Gaussian beam waist, $d$ is the distance to the nearest neighbor, and the maximum acceptable radial jitter, $r_0$, is chosen to be 150~nm. This value was chosen based on estimates of day to day misalignment observed when taking data in a 2D qubit array\cite{Xia2015,Maller2015a}. Equation (\ref{eq:wn}) yields the width, $w_n$, for each super Gaussian beam. All of the parameters used for the various addressing beams, as well as the crosstalk intensity values, are summarized in Table~\ref{tab:parameters}.

% \begin{table}%[H] add [H] placement to break table across pages
% \begin{ruledtabular}
% Lines of table here ending with \\
% \end{tabular}
% \end{ruledtabular}
% \end{table}

\begin{table}%[H] add [H] placement to break table across pages
%\begin{ruledtabular}
\begin{tabular}{c@{\hskip 0.25in} c@{\hskip 0.25in} c@{\hskip 0.1in}c}
\hline
\hline
Parameter & Value & $I / I_0$ & $I / I_0$ \\
 & & $\left(r = d\right)$ & $\left(r = d - r_0\right)$ \\
\hline
wavelength, $\lambda$ & 459 nm & & \\
trap spacing, $d$ & 3.8~$\mu$m & & \\
radial jitter, $r_0$ & 150 nm && \\
$w_{n = 2}$ & 2.30~$\mu$m & 0.0043 & 0.0065\\
$w_{n = 4}$ & 2.84~$\mu$m & 0.0017 & 0.0043\\
$w_{n = 6}$ & 3.09~$\mu$m & 0.0010 & 0.0043\\
$w_{n = 8}$ & 3.22~$\mu$m & 0.0006 & 0.0043\\
$w_{n = 10}$ & 3.30~$\mu$m & 0.0003 & 0.0043\\
\hline
\hline
\end{tabular}
%\setlength{\tabcolsep}{150pt}
%\end{ruledtabular}
\caption{List of parameters used for the various possible addressing laser beams. The width of the super Gaussian beams is chosen such that the crosstalk intensity, $I$, at a neighboring trap site is lower than, or equal to, that of a Gaussian beam even if the super Gaussian beam has a radial jitter up to a maximum allowable value of $r_0$.  $I_0$ is the intensity at the center of the focal plane of the Gaussian beam. \label{tab:parameters}}
\end{table}

\begin{figure}[!t]
 \centering
 \includegraphics[width=70mm]{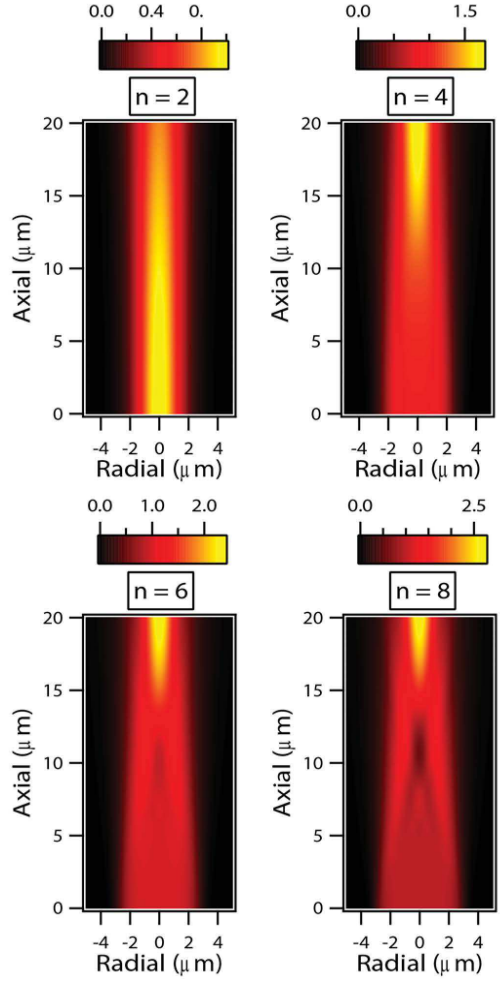}
\caption{(Color online) Image plots of the addressing beam intensity for a TEM$_{00}$ Gaussian beam, $n=$2, and super Gaussian beams of orders $n=$4, 6, and 8.  The parameters of the beams are given in Table~\ref{tab:parameters}. For each image plot, the focal plane is located at the bottom of the figure with the propagation direction towards the top of the page.}
\label{fig:2Dimages}
\end{figure}

\begin{figure}[!t]
 \centering
 \includegraphics[width=70mm]{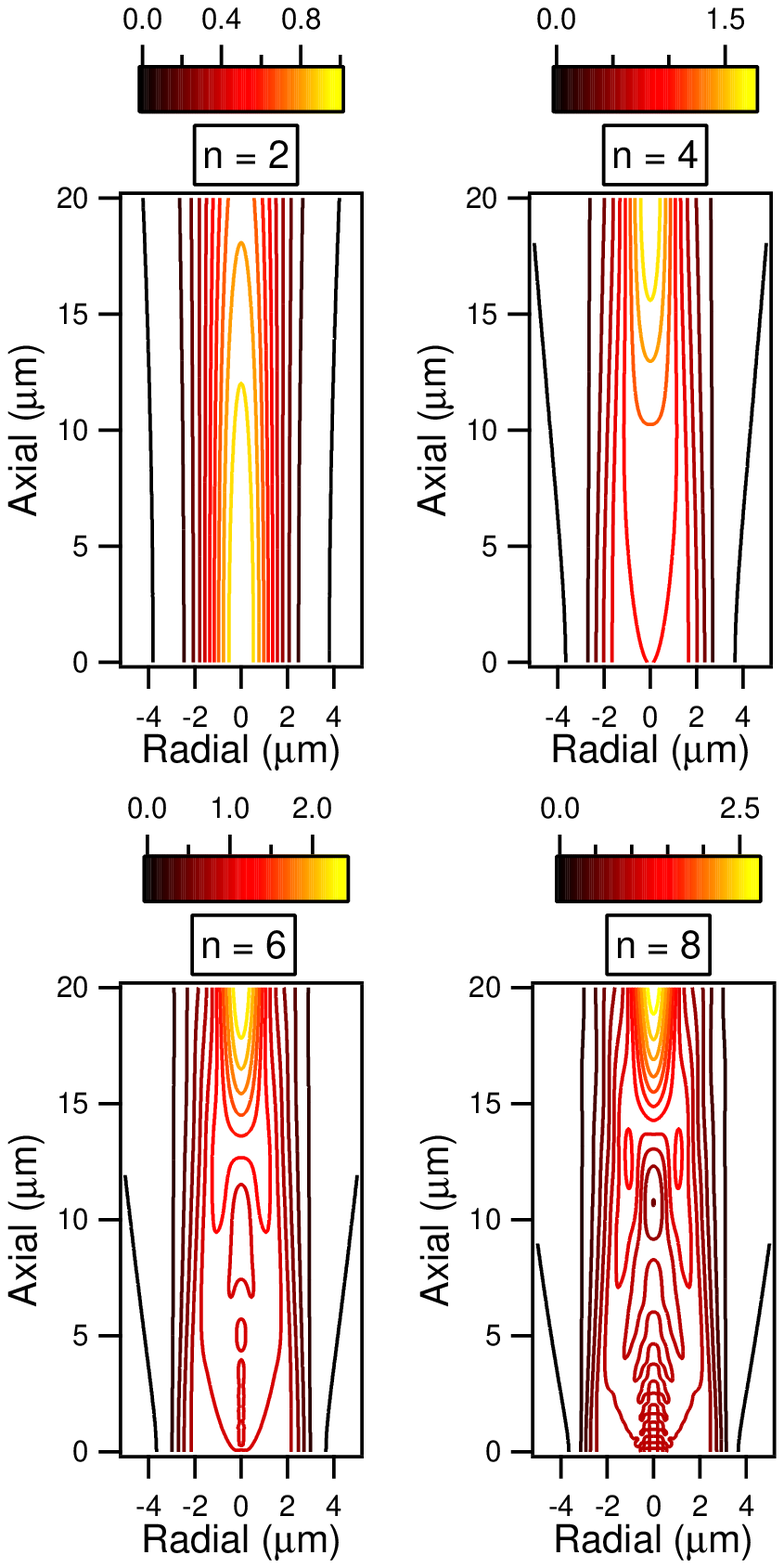}
\caption{(Color online) Contour plots of intensity distributions illustrated in Fig.~\ref{fig:2Dimages}.  The parameters of the beams are given in Table~\ref{tab:parameters}. For each plot, the lowest intensity contour line is for a normalized intensity of 0.0043;  i.e., the normalized intensity at a nearest neighbor trap site for a TEM$_{00}$ addressing laser beam.}
\label{fig:2Dcontour}
\end{figure}

Substituting Eq. (\ref{eq:SGE0}) into Eq. (\ref{eq:RSdiffintegral}) and integrating, we obtain the electric field of the super Gaussian beam at points beyond the input plane.    Figure~\ref{fig:2Dimages} is a collection of calculation results for the propagation of super Gaussian beams with various orders $n$, and a light wavelength of 459~nm.  Figure~\ref{fig:2Dcontour} is a collection of contour plots for the same super Gaussian beams depicted in Fig.~\ref{fig:2Dimages}.  For each of the figures the focal plane, $z = 0$, is located at the bottom of the plot and the beam propagation direction (increasing $z$ values) is towards the top of the page.

As observed in Fig.~\ref{fig:2Dimages}, the propagation of a super Gaussian beam is quite different than that of a Gaussian beam.  A few notable differences between a Gaussian beam, $n=2$, and a super Gaussian beam are apparent in variations in the radial intensity distributions as a function of location along the beam propagation direction, and merit some discussion.

\begin{figure}[!t]
 \centering
 \includegraphics[width=70mm]{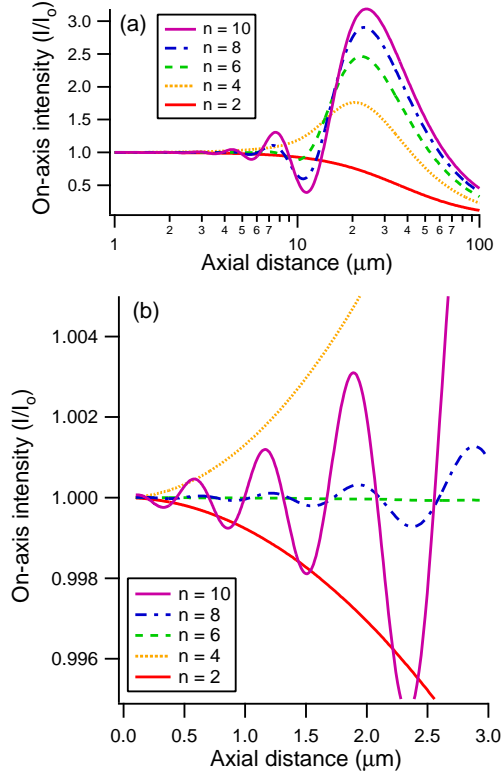}
\caption{(Color online) On-axis intensities for a TEM$_{00}$ Gaussian beam, $n=$2, and super Gaussian beams of orders $n=$4, 6, 8, and 10.  Part (a) illustrates the on-axis intensity behaviors over a large axial range out to $z = 100~\mu$m, and part (b) illustrates their behaviors near the focal plane.  For reference, the axial width of the atomic trapping volume extends to $\sigma_z \approx 1 - 2~\mu$m. The wavelength and beam waist widths for each value of $n$ are given in Table~\ref{tab:parameters}.}
\label{fig:GSGonaxis}
\end{figure}

The radial intensity profiles for both Gaussian and super Gaussian beams at the focal plane are illustrated in Fig.~\ref{fig:SGvsn}.  As mentioned previously, the radial beam profile for $n = 2$ has a Gaussian shape for all axial locations, and the highest intensity is found at the center of the beam in the focal plane. The radial beam profile for super Gaussian beams changes in a variety of different ways as a function of the axial location.  First, for all super Gaussian orders, as the beam propagates in the $+z$-direction the width of the central uniform intensity region narrows while the central intensity reaches a maximum value much higher than $I_0$ some distance away from the focal plane. Second, for super Gaussian orders greater than $n=4$ as the beam propagates away from $z=0$ oscillations in the intensity are observed in both the radial and axial directions.  These variations in the intensity are more apparent in the contour plots of Fig.~\ref{fig:2Dcontour}, and the on-axis intensity plots illustrated in Fig.~\ref{fig:GSGonaxis}.  As the order of the super Gaussian increases, so does the number and amplitude of axial and radial oscillations of the intensity.  Finally, the divergence of the beam for very low normalized intensities grows significantly as the order of the super Gaussian increases, as observed in the lowest normalized intensity value contour line for each plot in Fig.~\ref{fig:2Dcontour}.  

The lowest contour line for each plot in Fig.~\ref{fig:2Dcontour} has been manually set to a normalized intensity value of 0.0043.  This particular value for the lowest common contour line for all plots was chosen because that is the normalized intensity in the focal plane at the center of a neighboring trap site for the experimental Gaussian beam parameters modeled in this investigation, and discussed in Sec.~\ref{sec:Rabisimulation}. The normalized intensity at a neighboring trap site represents a measure of the crosstalk between an addressing beam for an atom in the trap site being addressed and another atom located in an adjacent trap.  The width used for each super Gaussian beam is such that the crosstalk normalized intensity value in the focal plane is the same for all super Gaussian orders if the beam has a maximal radial offset of $r_0$.  

The divergence of the beam can become a significant factor for either of two scenarios:  (1) if the axial size of the trapping volume overlaps with the divergence of an addressing beam for a neighboring trap site, or (2) if there is an axial misalignment, or jitter, between the focal plane of the addressing beam and the plane of the array of trap sites.  For reference, the axial confinement of atoms in these traps for the parameters used in Ref.~\cite{Piotrowicz2013} and this work, is $\sigma_z \approx 1 - 2 \mu$m.

\begin{figure}[!t]
\centering
\includegraphics[width=70mm]{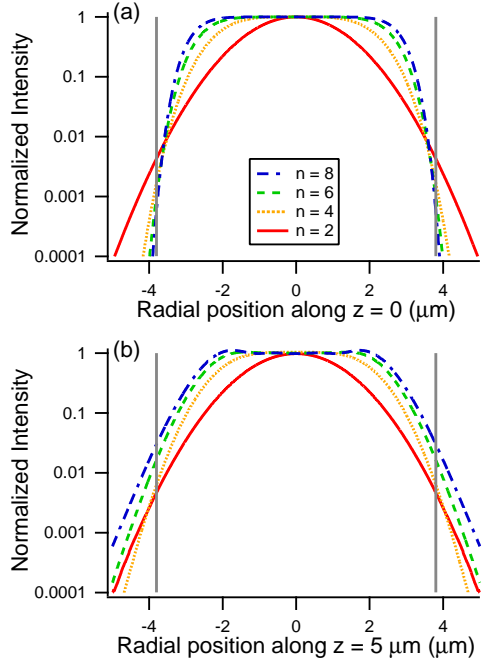}
\caption{(Color online)  Plots of the normalized radial intensity distribution of a 459-nm addressing laser beam as a function of radial distance from the center of the beam (a) in the focal plane, and (b) at an axial distance of 5~$\mu$m beyond the focal plane.  Vertical gray lines mark the location of the nearest neighbor trap site, or 
$r = 3.8~ \mu$m.  Beam parameters for these calculations are given in Table~\ref{tab:parameters}.}
\label{fig:GSGxtalk}%
\end{figure}

Figure~\ref{fig:GSGxtalk} illustrates the increasing beam divergence with increasing super Gaussian order.  Figure~\ref{fig:GSGxtalk}(a) is a plot of the radial intensity distribution at the focal plane, or $z=0$.  The vertical gray lines represent the location of a neighboring trap site located at a radial distance of $r = d$.  Note that all super Gaussian beams have a normalized intensity value at a neighboring trap site lower than that of a TEM$_{00}$ beam because we accounted for possible jitter up to $r_0$ in our choice for the widths of the super Gaussian beams.  Also note that for radial distances greater than $d$ the normalized intensity for all super Gaussian beams drops off faster with radial distance for higher values of $n$. Figure~\ref{fig:GSGxtalk}(b) is a plot of the radial intensity distributions at a distance of $z = 5~\mu$m from the focal plane.  By visually comparing (b) to (a) we can see that the normalized intensity value of super Gaussian beams at a radial distance of $d$, and beyond, is now significantly higher for increasing super Gaussian orders illustrating the greater divergence of super Gaussian beams as they propagate away from the focal plane.  

When choosing which type of beam to use to address the target atom an important side effect to continually monitor is the crosstalk intensity of the addressing beam on an atom in an adjacent trap site.  Beam misalignment, or jitter, of the addressing beam can negatively affect the state of an atom in a neighboring trap site due to the crosstalk intensity.   

\begin{figure}[!t]
\centering
\includegraphics[width=70mm]{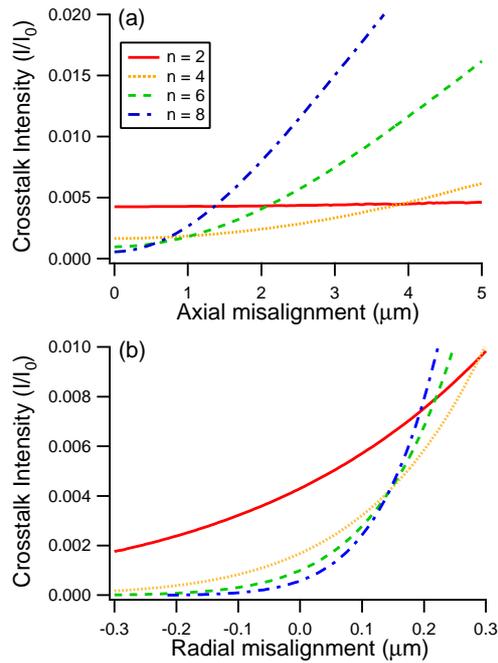}
\caption{(Color online)  Calculations for the crosstalk intensity at the location of a neighboring trap site at $d = 3.8~\mu$m (a) as a function of axial misalignment between the focal plane of the addressing beam and the plane containing the array of trap sites, and (b) as a function of radial misalignment of the addressing beam for the focal plane of the beam coplanar with the trap site array.  Beam parameters for these calculations are given in Table~\ref{tab:parameters}.}
\label{fig:GSGxtalkrandz}%
\end{figure}

To help illustrate how the crosstalk intensity depends upon the axial misalignment between the focal plane of the super Gaussian beam and the plane of the trap arrays, Fig.~\ref{fig:GSGxtalkrandz}(a) is a plot of the normalized intensity value as a function of axial distance from the focal plane for a fixed radial distance of $3.8 ~\mu$m from the center of the addressing beam.  The axial misalignment values where the crosstalk intensity for super Gaussian powers of $n = 8$, 6, and 4, is equal to that of a misaligned Gaussian beam are $z = 1.36 ~\mu$m, $2.10~\mu$m, and $3.90~\mu$m, respectively.  If the axial misalignment is smaller than these values, the crosstalk intensity at a neighboring trap site is lower for the respective super Gaussian beam orders than it is for the TEM$_{00}$ beam.  If the axial misalignment is greater than these values, the crosstalk intensity is lower for the TEM$_{00}$ than the respective Gaussian beam orders.

Figure~\ref{fig:GSGxtalkrandz}(b) illustrates the dependence of the crosstalk intensity at a neighboring trap site on radial misalignment of the addressing laser beam.  Here, the reasoning behind our choice for the widths of the super Gaussian beams, according to Eq. (\ref{eq:wn}), is visually illustrated.  As the order of the super Gaussian beam increases, the crosstalk intensity at a neighboring trap site changes much more rapidly than for a Gaussian beam.  The point at which the $n = 4$, 6, and 8 curves are all equal is for a radial misalignment equal to that of our chosen maximum acceptable jitter value, or $r = r_0 = 150$~nm.  If parameter $r_0$ is not included in Eq. (\ref{eq:wn}) (or set to zero) then the crosstalk intensity for all values of $n$ would be the same for an aligned addressing beam; i.e., the curves of Fig.~\ref{fig:GSGxtalkrandz}(b) would all cross at a radial misalignment of zero.  However, if any jitter was present (as is always the case with experimental laser beams) then the crosstalk intensity at a neighboring trap site would be significantly worse for any value of $n>2$.  By including the  maximum acceptable jitter parameter, $r_0$, into our choice for the width of each super Gaussian beam, the higher crosstalk intensity values for $n>2$ are pushed radially out beyond our chosen acceptable amount of jitter. The radial misalignment values where the crosstalk intensity for super Gaussian powers of $n = 8$, 6, and 4, is equal to that of the Gaussian beam are $r = 194$~nm, $212$~nm, and $287$~nm, respectively.

\subsection{Density weighted intensity variance calculations}
\label{sec:intvariance}

An atom confined within a Gaussian beam array trap and exposed to addressing beams will exist in a three-dimensional volume having a spatial variation in the intensity of the addressing and trapping beams.  Consequently, the atom will experience position-dependent differential AC Stark shifts and Rabi frequencies.  In this section, we computationally investigate the magnitude of the spatial intensity variations an atom at a particular temperature will experience by calculating the density weighted intensity variance described by Eq. (\ref{eq:ABIVgeneral}).  As discussed in Sec.~\ref{sec:intvar}, the more uniform the addressing beam intensity is over the volume of space occupied by the atom, the lower the value of the intensity variance.  If the volume occupied by the trapped atom is flooded with a uniform addressing beam intensity then the value of the intensity variance would be zero.

First, we will investigate the intensity variance for a perfectly aligned addressing beam; i.e., the beam waist is coplanar with the array of trap sites, and the addressing beam is colinear with the trapping axis.  Table~\ref{tab:variancealigned} is a collection of intensity variance values for these alignment conditions and an addressing laser beam with increasing values of the order $n$ of the super Gaussian.  All of the intensity variances reported in Table~\ref{tab:variancealigned} are for a trap spacing of 3.8~$\mu$m, a trap laser waist of 1.73~$\mu$m, and an atomic temperature of 20~$\mu$K, which yields, using Eqs. (\ref{eq:sigmaxyz}), a radial and axial confinement of $\sigma_x=$ 0.17~$\mu$m and $\sigma_z=$ 1.8~$\mu$m, respectively. The trap laser waist was chosen such that $s=\frac{d}{w_0}=2.197$, the optimum $s$ for the deepest trap from \cite{Piotrowicz2013}. The choices for the parameters for the Gaussian and super Gaussian addressing beam are discussed in Sec.~\ref{sec:SGcalcs} and given in Table~\ref{tab:parameters}.

\begin{table}%[width=86mm]%[H] add [H] placement to break table across pages
%\begin{ruledtabular}
\begin{tabular}{ r@{\hskip 0.25in} c@{\hskip 0.25in}  l}
\hline
\hline
$n$ & beam radius & $\sigma_{I \Psi}$ \\
 & ($\mu$m) & \\
\hline
2 & 2.30 & 0.01 \\
4 & 2.84 & 0.003 \\
6 & 3.09 & 0.00007 \\
8 & 3.22 & 0.0004 \\
10 & 3.30 & 0.002 \\
\hline
\hline
\end{tabular}
\setlength{\tabcolsep}{36pt}
%\end{ruledtabular}
\caption{Intensity variance values for a 20-$\mu$K atom in a Gaussian or super Gaussian beam with no misalignment between the trap site and the 459-nm wavelength addressing laser.  The value of each beam radius at the focal plane $z=0$ used in the calculations are given in Table~\ref{tab:parameters}, and repeated here for convenience.}
\label{tab:variancealigned}
\end{table}

As the value of $n$ increases from 2 to 4 to 6, we see that the intensity variance decreases due to the three-dimensional beam intensity distribution becoming more uniform over the volume of space occupied by the atom.  As the value of $n$ increases from 6 to 8 to 10 the intensity variance actually increases with an increase in the order of the super Gaussian.  Within the plane of the beam waist, the intensity distribution continues to become more uniform in the central region of the beam, as observed in Fig.~\ref{fig:SGvsn}.  However, it is the increase in the oscillations of the intensity distribution of the beam outside of the beam waist plane, as observed in Figs.~\ref{fig:2Dcontour} and~\ref{fig:GSGonaxis}, which force the intensity variance to increase as $n$ increases from 6 to 8 to 10.  Even for the limited axial range of the trapping volume (out to approximately $z = 1.8 ~\mu \rm m$ for a 20-$\mu$K atom) we can see in the contour plots in Fig.~\ref{fig:2Dcontour} that the number and frequency of intensity oscillations in the middle of the trapping volume noticeably increases from $n=6$ to $n=8$.

\begin{figure}[!t]
 \centering
 \includegraphics[width=70mm]{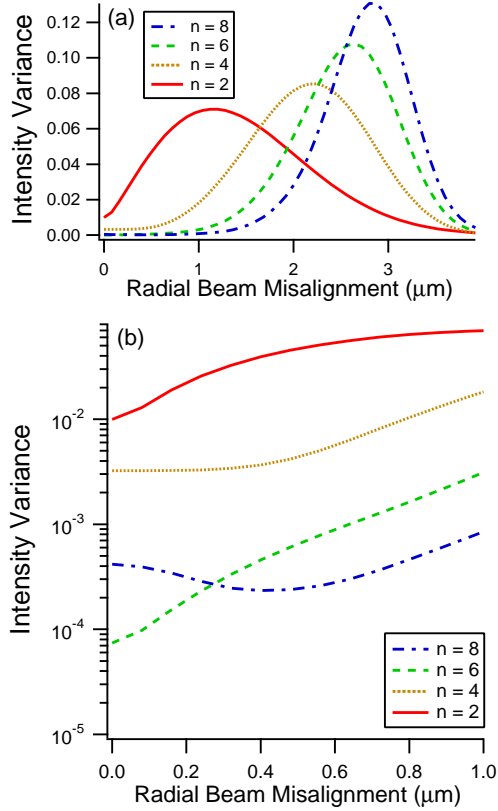}
\caption{(Color online) Calculated intensity variance for an atom with a temperature of 20~$\mu$K in a 459-nm wavelength TEM$_{00}$ Gaussian beam, $n=$2, and super Gaussian beams of orders $n=$4, 6, and 8 as a function of the radial misalignment between the addressing laser and the trap site.  Laser beam radii are the same as in Tables~\ref{tab:parameters} and~\ref{tab:variancealigned}.  Part (a) illustrates intensity variance values out to a radial misalignment equal to that of the location of the nearest neighbor trap site of $x = 3.8 \mu$m.  Part (b) illustrates the intensity variance on a log scale for a radial misalignment of less than one micron.}
\label{fig:variancevsx}
\end{figure}

Second, we investigate the intensity variance for a radial misalignment between the addressing beam and the trap site as a function of the order $n$ of the super Gaussian beam.  For these conditions, it is assumed that the volume occupied by the atom is axially and radially located at the center of the Gaussian beam array trap ($x=y=z=0$).  The addressing beam is assumed to be normal to the plane of the array (the $X$-$Y$ plane) with the beam waist plane coplanar with the trapping array (both located at $z=0$).  The addressing beam misalignment is only in the radial $x$-direction.  Fig.~\ref{fig:variancevsx} is a collection of calculation results of the intensity variance as a function of the radial misalignment between the addressing beam and the center of the volume occupied by the trapped atom for a TEM$_{00}$ Gaussian beam and super Gaussian beams of orders $n$ = 4, 6, and 8.

In Fig.~\ref{fig:variancevsx}(a) we see the behavior of the intensity variance out to a radial distance equal to that of the separation distance between trap sites, or $x=d=3.8~ \mu$m.  For every value of $n$ the intensity variance increases to a maximum value and then decreases again.  This behavior of the value of the intensity variance corresponds to the rate of change of the intensity beam profile of the addressing beam for each value of $n$.  The beam profiles of each value of $n$ are previously illustrated in Fig.~\ref{fig:SGvsn}.  For the $n=2$ TEM$_{00}$ Gaussian beam, the intensity immediately starts falling with the radial distance away from the center of the beam.  Consequently, the intensity variance immediately starts to increase even for a small radial misalignment of the addressing beam.   The maximum value of the intensity variance ($x \approx 1$~micron) also corresponds with the location of the highest rate of change, the inflection point, of the beam intensity profile.  The relatively uniform regions of intensity at the center of the super Gaussian beam profiles result in a much lower intensity variance for small radial misalignments ($x < 1~\mu$m).  Each of the higher super Gaussian exponent values of $n$ have a maximum intensity variance located at increasing values of the radial misalignment corresponding to the inflection point of the beam profile being located further away from the center of the beam.  Additionally, as observed in Fig.~\ref{fig:SGvsn}, higher values of $n$ result in steeper sides of the beam profile which results in an increasing value for the maximum intensity variance.  

Figure~\ref{fig:variancevsx}(b) shows the intensity variance on a logarithmic scale for small radial beam misalignments.  As the numbers in Table~\ref{tab:variancealigned} reveal, the intensity variance for no misalignment between the addressing beam and the trap site significantly decreases as $n$ increases from 2 to 4 to 6.  Then, due to the small on-axis intensity oscillations just beyond the beam waist for high values of $n$, the intensity variance increases from $n=6$ to $n=8$.  These on-axis intensity oscillations near the beam waist plane for $n=8$ are observed in Fig.~\ref{fig:GSGonaxis}(b).  As the $n=8$ addressing beam is radially misaligned the trapping volume walks off of the on-axis intensity oscillations and into a region of more uniform intensity located between the optical axis and the wall of the beam.  Hence, the intensity variance of the $n=8$ beam initially decreases for increasing radial misalignment.  For perfect trap-addressing beam alignment, or a radial misalignment less than 0.25 microns, the $n=6$ super Gaussian beam provides the lowest intensity variance. 

\begin{figure}[!t]
 \centering
 \includegraphics[width=70mm]{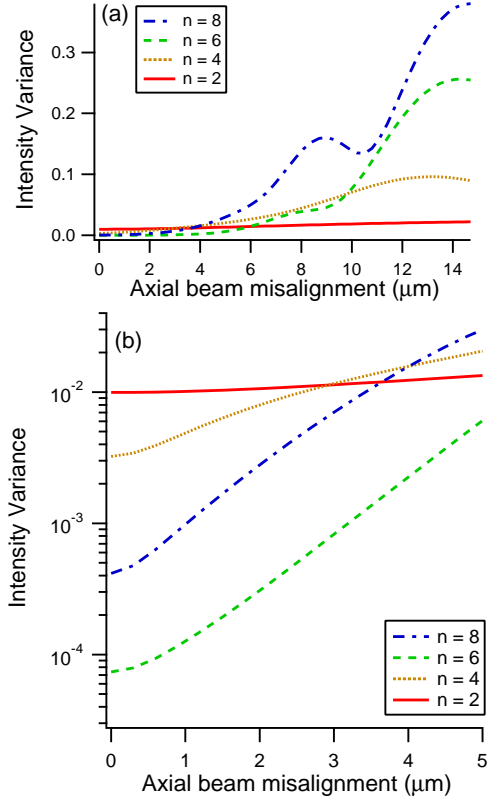}
\caption{(Color online) Calculated intensity variance for an atom with a temperature of 20~$\mu$K in a 459-nm TEM$_{00}$ Gaussian beam, $n=$2, and super Gaussian beams of orders $n=$4, 6, and 8 as a function of the axial misalignment between the beam waist of the addressing laser and the center of the trap site.  Laser beam radii are the same as in Tables~\ref{tab:parameters} and~\ref{tab:variancealigned}.}
\label{fig:variancevsz}
\end{figure}

Finally, we  investigate the intensity variance for an axial misalignment between the beam waist plane and the axial center of the trap site as a function of the order $n$ of the super Gaussian beam.  For these conditions it is assumed that the volume of space occupied by the trapped atom is centered around $x=y=z=0$, and the propagation of the addressing beam is along the $z$-axis; i.e., along the radial center of the trapping volume.  However the beam waist plane of the addressing beam is not coplanar with the trapping array, $z=0$, but rather located at some other axial location where $z > 0$.  Figure~\ref{fig:variancevsz} is a collection of calculation results for the intensity variance for values of $n$ from 2 to 8 as a function of the axial misalignment between the trapping volume and the beam waist plane of the addressing beam. Figure~\ref{fig:variancevsz}(a) illustrates the behavior of the intensity variance over a large range of axial misalignment values, and Fig.~\ref{fig:variancevsz}(b) shows an expanded view of the intensity variances for an axial misalignment of less than 5~microns.  Figure~\ref{fig:variancevsz}(b) reveals that for all values of $n$, and small to moderate axial misalignment, the intensity variance increases monotonically with an increase in the misalignment of the addressing beam. The intensity variance for $n=8$ is higher than that for $n=6$ due to the increase in the oscillations of the intensity along the central region of the addressing beam.  The central intensity oscillations are observed in Fig.~\ref{fig:GSGonaxis}(b). Overall, a super Gaussian order of $n=6$ provides the lowest intensity variance for all axial misalignment values less than 5~microns, and radial misalignment values less than 0.25~microns, due to the relatively uniform intensity over this volume of space, as can be observed in Figs.~\ref{fig:2Dimages} and~\ref{fig:2Dcontour}.   

\begin{table}%[width=86mm]%[H] add [H] placement to break table across pages
%\begin{ruledtabular}
\begin{tabular}{ r@{\hskip 0.25in} c@{\hskip 0.25in}  l}
\hline
\hline
$n$ & beam radius & $\sigma_{I \Psi}$ \\
 & ($\mu$m) & \\
\hline
5.5 & 3.04 & 0.00020 \\
6 & 3.09 & 0.00007 \\
6.5 & 3.13 & 0.00006 \\
7 & 3.16 & 0.00008 \\
7.5 & 3.19 & 0.00013 \\
\hline
\hline
\end{tabular}
\setlength{\tabcolsep}{36pt}
%\end{ruledtabular}
\caption{Intensity variance values of a 20-$\mu$K atom in a super Gaussian addressing beam with no misalignment between the trap site and the 459-nm wavelength addressing laser for various super Gaussian power values between $n = 5.5$ and 7.5.}
\label{tab:variancealignednear6}
\end{table}

For even values of the super Gaussian power, the minimum intensity variance for addressing beams having an axial misalignment of 6~$\mu$m or less occurs for a value of $n=6$.  Experimentally created super Gaussian beams may have a focal plane intensity distribution which varies from that of the desired theoretical beam profile.  Therefore, we also investigate the intensity variance for a variety of super Gaussian powers around $n = 6$.  The resulting intensity variances for $n$ values ranging from 5.5 to 7.5 are reported in Table~\ref{tab:variancealignednear6}. We find that the minimum intensity variance occurs for a super Gaussian power of $n=6.5$.  Variations of $\Delta n = 0.5$ result in small changes ($\leq$ 33\%) in the intensity variance, whereas $\Delta n = 1$ results in a factor of 2 or 3 increase in the intensity variance.  We conclude that some level of deviation from the ideal beam shape is acceptable in an experimental setup.  Note that the simulations presented in the next section are performed for $n=6$, which is a slightly less than ideal case, meaning we could achieve even better results for $n=6.5$ than are presented here.

\subsection{Simulation of Rabi oscillations}
\label{sec:Rabisimulation}

In this section, we investigate what type of intensity profile is best suited for the addressing laser beams used to drive single qubit rotations. We use a pair of Raman laser beams with atoms trapped in the Gaussian beam array described in Sec.~\ref{sec:theorysetup}. The experimental challenges include crosstalk with atoms at neighboring trap sites and shifts in beam alignment resulting in changes of the differential AC Stark shifts and Rabi frequencies experienced by the atomic qubits. Any unintended changes in these results in incorrect laser pulse times, incomplete population transfer, and thus gate errors. We compare the Rabi oscillations generated by Gaussian and super Gaussian beams to assess which order $n$ yields the most consistent Rabi oscillation amplitudes and frequencies.

We simulated Rabi oscillations between the hyperfine ground states $6^2S_{1/2},F=3$ and $6^2S_{1/2}, F=4$ of $^{133}$Cs driven by Raman transitions via the $7^2P_{1/2}$ level.  The two Raman laser beams used in this simulation have a wavelength of 459~nm with a detuning of 20~GHz above the $6^2S_{1/2}\rightarrow7^2P_{1/2},$ transition. For the simulation of Gaussian addressing beams, we used a beam waist of 2.3~$\mu$m and a power of 5~$\mu$W per beam at the atoms. We performed a Monte Carlo simulation of the motion and atomic state evolution of an atom in the Gaussian beam array trap. The motion was determined by solving the classical equation $\frac{d^2 {\bf r}_a}{dt^2}={\bf F}/m_a$ with ${\bf r}_a$ the atomic position and ${\bf F}=-\nabla U_T$ the gradient force from the trap potential  as given by Eq. (\ref{eq:trappotential}).  For typical experimental parameters the contribution to the force from the Raman light is negligible and will be neglected.   We ran the simulation for each of our scenarios for 100 atoms, each placed in the trap according to a Gaussian spatial distribution using Eqs. (\ref{eq:sigmaxyz}) and a Maxwellian velocity distribution in three dimensions for the given atom temperature. The motion was simulated 
with  time steps of 1~$\mu$s up to a total time of 150~$\mu$s. The time steps were chosen to be  much smaller than the trap oscillation periods of the atoms, since the radial trap frequency is $\omega_x=2\pi\times$31.5~kHz and the axial trap frequency is $\omega_z=2\pi\times$3.2~kHz for the trap laser parameters we used (see Sec.~\ref{sec:theorysetup}). Based on the position of each atom at each time step, we then calculated the differential AC Stark shift and on-resonance Rabi frequency at this position with Eqs. (\ref{eq:diffStarkRaman}, \ref{eq:Rabifrequency}) and used Eq. (\ref{eq:statevector}) to evolve the state of each atom. We tuned the Raman laser pair to the $F=3$ and $F=4$ hyperfine ground state splitting with AC Stark shifts for an atom located at the center of the trap and perfectly aligned addressing beams. Thus, $\Delta$ from Eq. (\ref{eq:statevector}) is the change in differential AC Stark shift from that of an atom at the center. At this position, the differential AC Stark shift due to the Raman laser beams is -13.49~kHz, and the on-resonance Rabi frequency is 23.4~kHz.
For each scenario we investigated, we recorded the differential AC Stark shift due to the Raman laser beams, the effective on-resonance ($\Omega$) and off-resonance ($\Omega'$) Rabi oscillation frequencies, the 1/e Rabi oscillation amplitude decay time, $t_a$, and the $F=4$ population, $|c_e\left(t\right)|^2$, for $\pi$ and 3$\pi$ Raman pulse times. This allows us to compare the effects of atom temperature and laser beam misalignment on the consistency of the resulting quantum operations (e.g. $\pi$ pulses) for different laser beam intensity profiles.
For the differential AC Stark shift and the on-resonance Rabi frequency, we report the average value for 100 atoms at their initial positions. Their dependence on the laser intensity is identical (see Eqs. (\ref{eq:diffStarkRaman}, \ref{eq:Rabifrequency})), so a larger on-resonance Rabi frequency is correlated with a larger (more negative) differential Stark shift. The off-resonance Rabi frequency and Rabi oscillation amplitude decay time were determined by fitting a curve of the form
\begin{equation}
|c_e\left(t\right)|^2= A \sin^2\left(\frac{\Omega' t}{2}\right)e^{-t/t_a}+ B \left(1-e^{-t/t_b}\right)
\nonumber%\label{eq:fitfunction}
\end{equation}
to the $|c_e\left(t\right)|^2$ data, where $A$, $B$, $\Omega'$, $t_a$, and $t_b$ are the five fit parameters. $\Omega'$ and $t_a$ from this fit are reported as off-resonance Rabi frequency and 1/e Rabi oscillation amplitude decay time here. Because the decay time of the Rabi oscillations was very sensitive to small variations in atom positions, we ran each scenario ten times, and  reported the mean and standard deviation of the results. Statistical outliers based on the decay time (in units of $\pi$ pulse time) were removed from the analysis using the modified Thompson tau method \cite{Wheeler1996}.

\begin{figure}[!t]
 \centering
 \includegraphics[width=70mm]{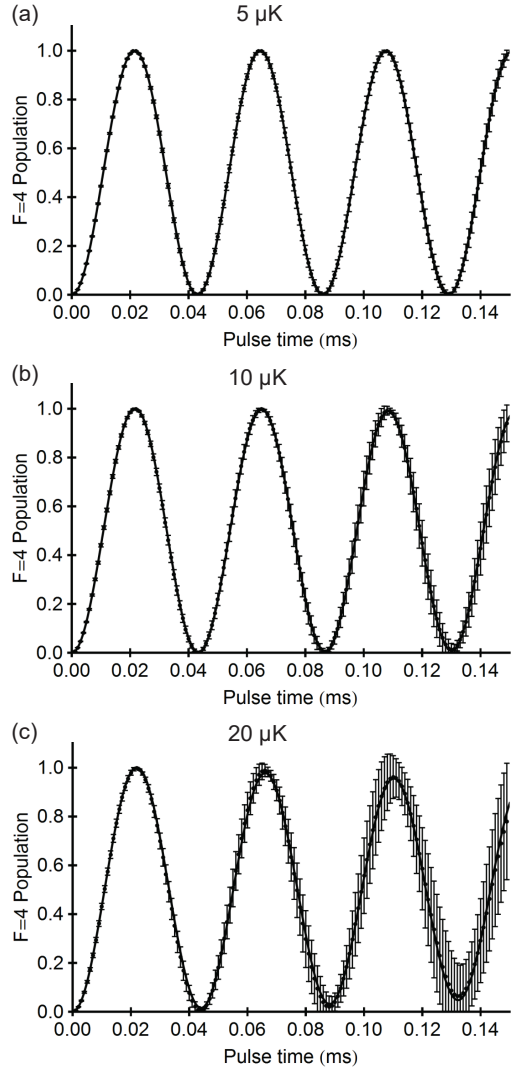}
\caption{$F=4$ hyperfine ground state population for Rabi oscillation between $F=3$ and $F=4$ states for Gaussian Raman laser beams for atom temperatures of (a) 5~$\mu$K, (b) 10~$\mu$K, (c) 20~$\mu$K.}
\label{fig.Raman_temperature}
\end{figure}

We first investigated the atom-temperature dependence of the Rabi oscillations. Figure \ref{fig.Raman_temperature} shows the resulting oscillations in a pair of Gaussian Raman laser beams for atom temperatures of 5~$\mu$K, 10~$\mu$K, and 20~$\mu$K, respectively. The data points shown are the average $F=4$ hyperfine ground state population for all 100 atoms, and the error bars indicate $\pm\sigma$ variations for each Raman laser pulse time. The data shown in the graphs is one representative sample out of ten repetitions of the simulation of 100 atoms. The numerical results are listed in Table~\ref{tab:RabiTdepGSG}.

\begin{table}%[width=86mm]%[H] add [H] placement to break table across pages
%\begin{ruledtabular}
\begin{tabular}{c  c  c  c  c  c  c  c  c}
\hline
\hline
$n$ & $T$ & $\frac{\Delta_{acR}}{2\pi}$ & $\frac{\Omega}{2\pi}$ & $\frac{\Omega'}{2\pi}$ & $t_a$ & $t_a$ & $\pi$ pop. & 3$\pi$ pop.  \\
 & ($\mu$K) & (kHz) & (kHz) & (kHz) & (ms) & ($\pi$ pulse time $\times$100) & ($\%$) & ($\%$) \\
\hline
2 & 5 & -13.398(9) & 23.233(15) & 23.23(2) & 22(6) & 10(3) & 99.863(4) & 99.839(9) \\
2 & 10 & -13.314(12) & 23.09(2) & 23.04(2) & 4.7(7) & 2.1(3) & 99.892(4) & 99.78(4) \\
2 & 20 & -13.11(3) & 22.73(5) & 22.64(6) & 1.12(14) & 0.51(6) & 99.81(2) & 99.0(2) \\
4 & 20 & -13.553(7) & 23.502(12) & 23.48(1) & 24(4) & 11(2) & 99.943(3) & 99.92(2) \\
6 & 20 & -13.4939(3) & 23.3994(5) & 23.388(2) & 67(7) & 32(3) & 99.9162(13) & 99.979(3) \\
8 & 20 & -13.493(2) & 23.397(3) & 23.384(2) & 58(10) & 28(5) & 99.913(2) & 99.976(4) \\
\hline
\hline
\end{tabular}
\setlength{\tabcolsep}{36pt}
%\end{ruledtabular}
\caption{Rabi oscillation results for atoms of various temperatures $T$ in Gaussian ($n=2$) and 20-$\mu$K atoms in super Gaussian ($n=4,6,8$) 459-nm wavelength addressing laser beams that are aligned with the trap site. The parameters listed are the differential AC Stark shift $\Delta_{acR}$ due to the Raman lasers, the effective on-resonance ($\Omega$) and off-resonance ($\Omega'$) Rabi frequencies, the 1/e Rabi oscillation amplitude decay time $t_a$, and the $F=4$ population for $\pi$ and 3$\pi$ Raman pulse times, respectively. For an atom at the center of the trap and the addressing laser beams, $\Delta_{acR}=-2\pi\times13.49$~kHz and $\Omega=2\pi\times23.40$~kHz. The average contributions to the differential AC Stark shift due to the trap laser are $\Delta_{acT}=2\pi\times0.684(4)$~kHz for a 5-$\mu$K atom, $2\pi\times$0.722(7)~kHz for a 10-$\mu$K atom, and $2\pi\times$0.809(6)~kHz for a 20-$\mu$K atom.}
\label{tab:RabiTdepGSG}
\end{table}

We find that in this temperature range, the differential AC Stark shift due to the Raman laser beams and the effective on- and off-resonance Rabi frequencies drop by a few percent from their 5-$\mu$K values (which are within 1\% of those for an atom at the center). The average contributions to the differential AC Stark shift due to the trap laser are $\Delta_{acT}=2\pi\times0.684(4)$~kHz for a 5-$\mu$K atom, $2\pi\times$0.722(7)~kHz for a 10-$\mu$K atom, and $2\pi\times$0.809(6)~kHz for a 20-$\mu$K atom. The $\pi$ and $3\pi$ Raman pulse time populations are within 1 percent of the 5-$\mu$K value, with the $3\pi$ populations being lower for larger temperatures. The decay time $t_a$ is approximately a factor of 20 shorter for 20~$\mu$K than for 5~$\mu$K, but it is still many times the duration of a $\pi$ pulse. Fig.~\ref{fig.Raman_temperature} shows that the spread of the $F=4$ population increases significantly (by a factor of five to ten) with temperature.

The higher the atom temperature, the faster and farther the atoms move in the trap. Farther away from the trap center on average, atoms sample a lower intensity of the Gaussian beam, leading to the observed reductions in differential AC Stark shift and Rabi frequencies (see Eqs.~\ref{eq:diffStarkRaman} and \ref{eq:Rabifrequency}). Because the atoms move farther at higher temperatures, they also sample a larger range of intensities, leading to a larger range of differential Stark shifts and Rabi frequencies. This effectively results in a range of Rabi oscillation amplitudes (see Eq. (\ref{eq:populations})), causing the spread of $F=4$ population observed in Fig.~\ref{fig.Raman_temperature}. Averaging the oscillations of 100 atoms with varying off-resonance Rabi frequencies results in the observed decay of the amplitude of the averaged oscillation. Thus, higher temperature atoms, which have the higher range of Rabi frequencies, have shorter decay times of their average Rabi oscillation.

Next, we investigated how the use of super Gaussian addressing laser beams affects the Rabi oscillations. We repeated the simulations for an atom temperature of 20~$\mu$K for super Gaussian beams of orders $n=4,6$, and $8$. The beam waists used are listed in Table~\ref{tab:parameters} and were chosen such that a radial beam displacement of 150 nm would result in the same intensity at a neighboring trap site as an aligned Gaussian beam with the parameters described above. The intensity at the center of each beam was set to match that of the Gaussian case, thus requiring more power per beam than the Gaussian beams. 

As shown in Table~\ref{tab:RabiTdepGSG}, the differential AC Stark shift due to the Raman laser beams and the effective on- and off-resonance Rabi frequencies for super Gaussian beams are within a fraction of a percent of the values for an atom at the center of the trap and addressing beams. They are lowest for the Gaussian beam ($n=2$), highest for the $n=4$ super Gaussian beam, and then lower and lower for $n=6$ and $n=8$, with $n=6$ having the values closest to those for an atom at the center of the trap and addressing beams. The Rabi oscillation amplitude decay times and the $\pi$ and $3\pi$ populations are higher for super Gaussians than those for a Gaussian beam and are lower the farther off their Stark shift and Rabi frequencies are from their center values. Thus, they are highest for the $n=6$ super Gaussian laser beam. The decay times are much improved (by a factor of approximately 20 to 60) when using super Gaussian beams. For $n=4$ the decay time is as long as that of a 5-$\mu$K atom sample in a Gaussian beam, and the decay times for $n=6$ and $n=8$ are about three times larger.

These results are consistent with those from Secs.~\ref{sec:SGcalcs} and \ref{sec:intvariance}. While the intensity of a Gaussian beam drops off in all directions, that of an $n=4$ super Gaussian increases away from the focal plane, leading to the observed increase in differential Stark shift and Rabi frequencies. Super Gaussian beams of order $n=6$ have a flat intensity profile near the trap center along the axial direction, making this the most constant intensity profile in all three dimensions, and thus having differential Stark shift and Rabi frequency values closest to those at the center of the trap. The $n=8$ super Gaussian has an intensity profile that oscillates around the focal plane value along the axial direction, on average leading to differential AC Stark shifts and Rabi frequencies very close to the center ones. The Rabi oscillation decay times match the trends of the intensity variance shown in Table~\ref{tab:variancealigned}, with higher intensity variance correlated with shorter decay times. Because $n=6$ has the lowest intensity variance, the differential AC Stark shift changes the least as the atom moves around the trap, leaving $\Delta$ close to zero and resulting in a smaller spread in Rabi oscillation amplitude and frequency, as observed by the longer Rabi oscillation decay times. Any experimental setup has a limit to the coldest atom temperature that can be achieved. Within these constraints, our results suggest that switching to a super Gaussian beam such as $n=6$ can much improve the consistency of the resulting Rabi oscillations (less spread, longer decay times) and thus quantum operations.

\begin{figure}[!t]
 \centering
 \includegraphics[width=86mm]{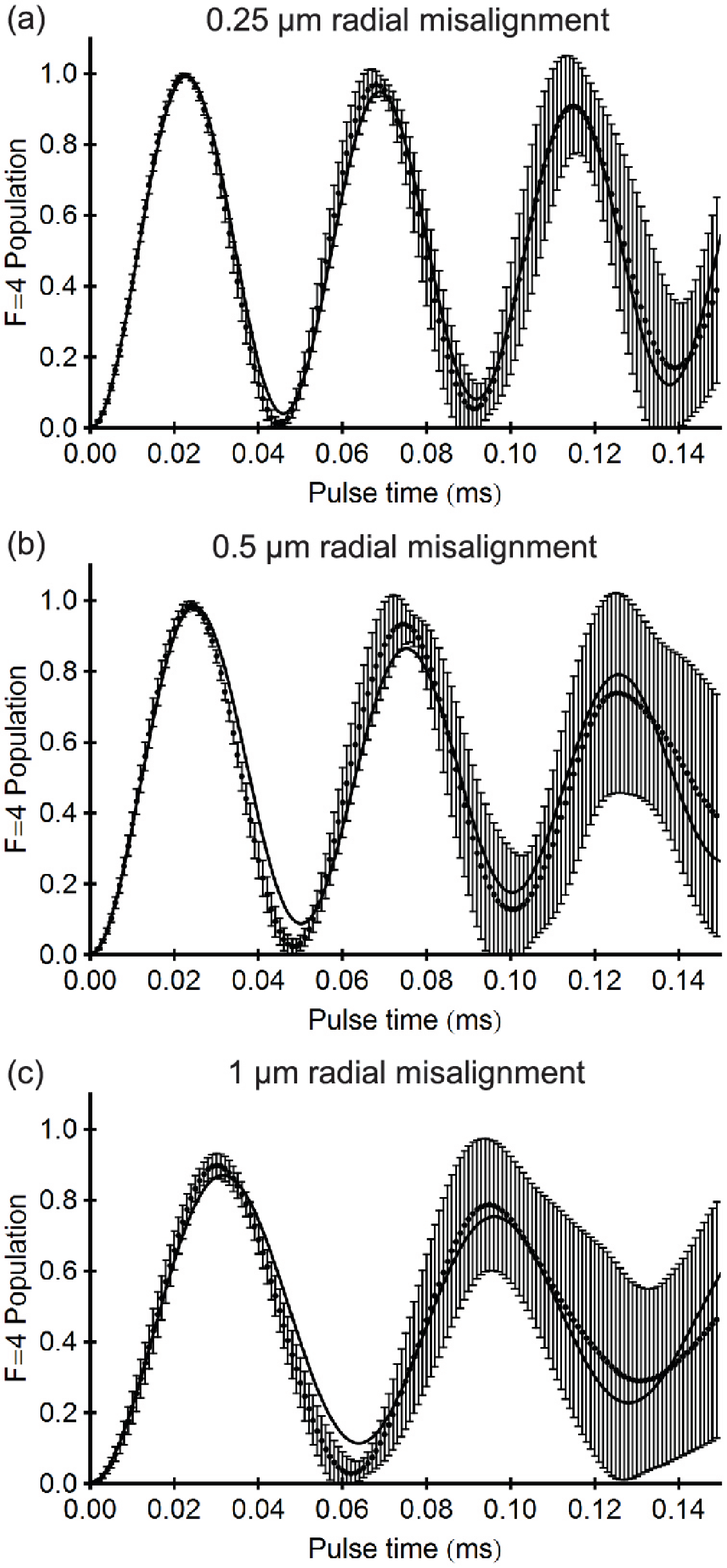}
\caption{$F=4$ hyperfine ground state population for Rabi oscillations between $F=3$ and $F=4$ states for Gaussian Raman laser beams for an atom temperature of 20~$\mu$K and a radial Raman beam misalignment of (a) 0.25~$\mu$m, (b) 0.5~$\mu$m, (c) 1~$\mu$m.}
\label{fig.Raman_misalign}
\end{figure}

In any experiment that requires addressing of a single trapped atom, addressing laser beam alignment is critical. To investigate the dependence of the Rabi oscillations  on beam alignment, we repeated the simulation for a radial Raman beam misalignment from the trap center of 0.25~$\mu$m, 0.5~$\mu$m, and 1~$\mu$m for an atom temperature of 20~$\mu$K for a Gaussian ($n=2$) addressing beam pair. Because the Raman beams share the same optical path, we only investigated a common shift of both Raman laser beams. The resulting time-dependent Raman $F=4$ populations are shown in Fig. \ref{fig.Raman_misalign}. The corresponding numerical results are shown in Table~\ref{tab:RabimisalignradialGSG}.

\begin{table}%[width=86mm]%[H] add [H] placement to break table across pages
%\begin{ruledtabular}
\begin{tabular}{c  c  c  c  c  c  c  c  c}
\hline
\hline
$n$ & $\Delta x$ & $\frac{\Delta_{acR}}{2\pi}$ & $\frac{\Omega}{2\pi}$ & $\frac{\Omega'}{2\pi}$ & $t_a$ & $t_a$ & $\pi$ pop. & 3$\pi$ pop.  \\
 & ($\mu$m) & (kHz) & (kHz) & (kHz) & (ms) & ($\pi$ pulse time $\times$100) & ($\%$) & ($\%$) \\
\hline
2 & 0.25 & -12.83(5) & 22.25(8) & 21.79(6) & 0.50(5) & 0.23(2) & 99.54(5) & 97.0(3) \\
2 & 0.5 & -12.01(9) & 20.8(2) & 19.9(1) & 0.204(15) & 0.085(6) & 98.69(8) & 93.8(7) \\
2 & 1 & -9.1(2) & 15.8(3) & 15.55(8) & 0.193(13) & 0.064(4) & 89.1(5) & 77.8(1.3) \\
4 & 1 & -13.03(3) & 22.60(6) & 22.26(3) & 1.1(2) & 0.50(8) & 99.79(3) & 98.6(2) \\
6 & 1 & -13.486(7) & 23.385(12) & 23.361(8) & 18(4) & 8.4(1.9) & 99.904(3) & 99.91(2) \\
8 & 1 & -13.498(3) & 23.406(6) & 23.394(4) & 45(7) & 22(3) & 99.919(2) & 99.975(9) \\
\hline
\hline
\end{tabular}
\setlength{\tabcolsep}{36pt}
%\end{ruledtabular}
\caption{Rabi oscillation results for atoms with a temperature of 20~$\mu$K in Gaussian beams ($n=2$) with varying amounts of radial misalignment $\Delta x$ between the trap site and the focus of the 459-nm wavelength addressing laser beams, and super Gaussian beams ($n=4,6,8$) with a radial misalignment of 1~$\mu$m. The parameters listed are the differential AC Stark shift $\Delta_{acR}$ due to the Raman lasers, the effective on-resonance ($\Omega$) and off-resonance ($\Omega'$) Rabi frequencies, the 1/e Rabi oscillation amplitude decay time $t_a$, and the $F=4$ population for $\pi$ and 3$\pi$ Raman pulse times, respectively. For an atom at the center of the trap and the addressing laser beams $\Delta_{acR}=-2\pi\times13.49$~kHz and $\Omega=2\pi\times23.40$~kHz. The average contribution to the differential AC Stark shift due to the trap laser is $\Delta_{acT}=2\pi\times0.809(6)$~kHz.}
\label{tab:RabimisalignradialGSG}
\end{table}

We see that radial beam misalignment has significant effects on all of the quantities we calculated. The differential Stark shift and the on- and off-resonance Rabi frequencies drop to about two thirds of their perfectly aligned values at 1~$\mu$m radial beam misalignment. Misalignment of 0.25~$\mu$m reduces the decay time by a factor of two, and misalignment of 1~$\mu$m reduces decay times to only a few multiples of the $\pi$ pulse time. Consequently, the $F=4$ population at $\pi$ and 3$\pi$ pulse times are significantly reduced, dropping as low as 89\% and 78\%, respectively, for 1~$\mu$m of radial misalignment. Furthermore, we see that the spread in $F=4$ population reaches values of about 33$\%$ at 0.15~ms when misaligning the Gaussian addressing beams radially.

These results are consistent with those of Secs.~\ref{sec:SGcalcs} and \ref{sec:intvariance}. For a Gaussian ($n=2$) beam, as the radial misalignment is increased, the intensity sampled by the atoms drops at the edge of the Gaussian beam (see Fig.~\ref{fig:SGvsn}), leading to a drop in differential AC Stark shift and Rabi frequencies. This means that if the Raman beams shift at this level after the pulse times have been carefully calibrated, the pulse times will be incorrect and the quantum operation performed will be different than intended. Also, with radial misalignment the intensity variance of the Gaussian beam increases (see Fig.~\ref{fig:variancevsx}), since its intensity changes more rapidly as we reach its inflection point around 1~$\mu$m. The atoms thus sample a larger range of intensities, resulting in a larger range of differential AC Stark shifts and thus Rabi oscillation amplitudes and frequencies. This leads to the spread in $F=4$ populations seen in Fig.~\ref{fig.Raman_misalign} and faster decay. 

\begin{figure}[!t]
 \centering
 \includegraphics[width=86mm]{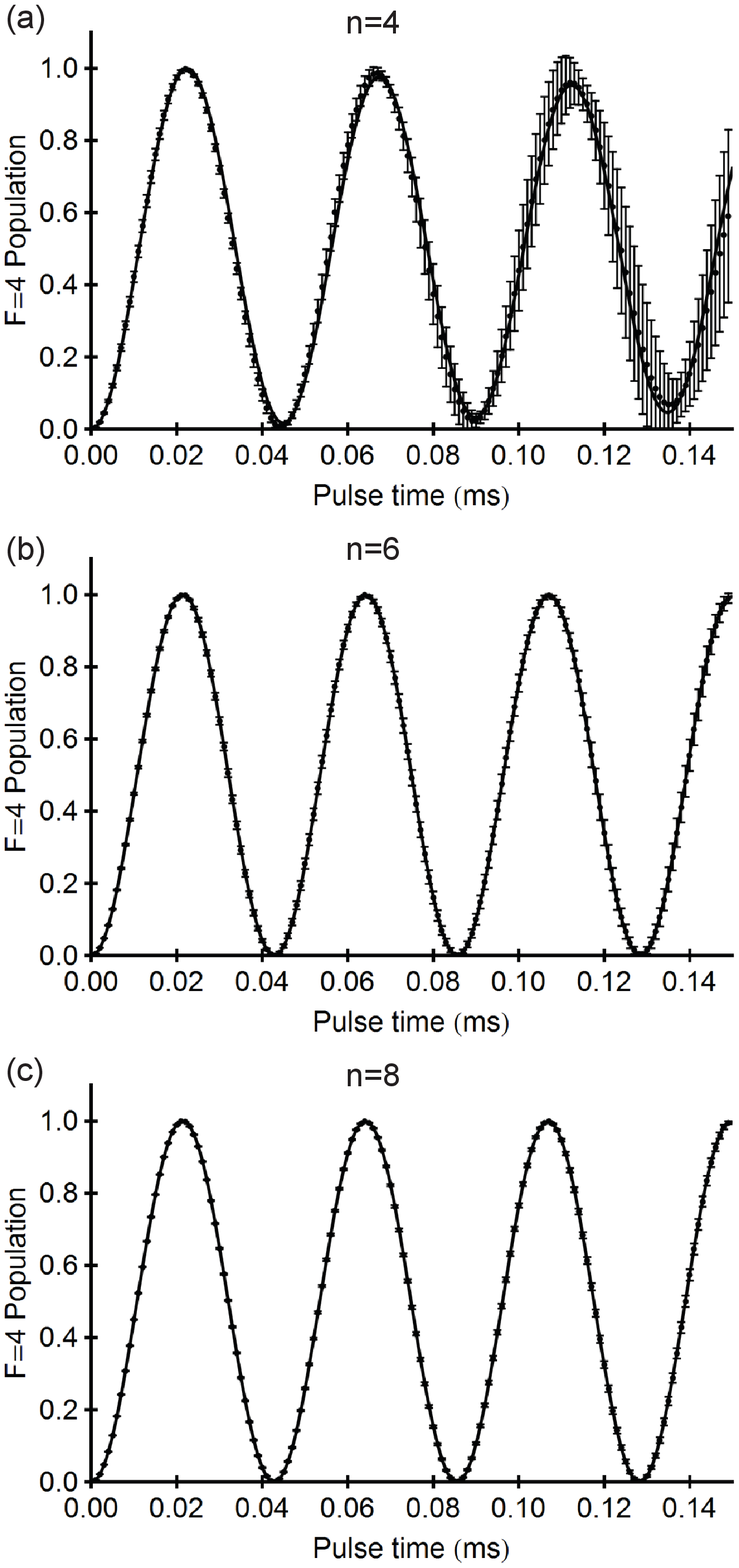}
\caption{$F=4$ hyperfine ground state population for Raman transition between $F=3$ and $F=4$ states for super Gaussian Raman laser beams for an atom temperature of 20~$\mu$K and a radial Raman beam misalignment of 1~$\mu$m. (a) $n=4$, (b) $n=6$, (c) $n=8$.}
\label{fig.Raman_superGaussian}
\end{figure}

Again, we investigated how super Gaussian addressing laser beams performed in this scenario. The results are shown in Fig.~\ref{fig.Raman_superGaussian} and Table~\ref{tab:RabimisalignradialGSG}. In each case, we assumed an atom temperature of 20~$\mu$K and a radial beam misalignment of 1~$\mu$m. The amount of misalignment encountered in the laboratory is determined by the limitations of the experimental setup. We therefore assumed the same amount of misalignment for all super Gaussian orders, regardless of beam size.  We find that the higher the order $n$ of super Gaussian used, the closer the differential Stark shift and Rabi frequencies, and the $\pi$ and 3$\pi$ $F=4$ populations get to those for perfect radial alignment, deviating only a few hundredths of percent or less from the values at the center of the trap. Similarly, the higher the order $n$, the longer the decay times. For $n=4$, the decay time is more than double that of a Gaussian beam that is only misaligned by 0.25~$\mu$m, and that for $n=8$ is almost 100 times as large. While for $n=4$ the spread in $F=4$ population is comparable to that of the aligned Gaussian beam case, the spread for $n=6$ is less than that of a $5~\mu$K atom in an aligned Gaussian laser beam, and $n=8$ has an even smaller spread than that. 

This is consistent with Secs.~\ref{sec:SGcalcs} and \ref{sec:intvariance}. 
Figure~\ref{fig:SGvsn} shows that the higher the super Gaussian order $n$, the flatter the intensity profile remains around the $1~\mu$m range examined here, making all parameters closer to those for an atom at the center of the trap for aligned addressing beams. Figure~\ref{fig:variancevsx} shows that at 1~$\mu$m radial misalignment the intensity variance is smaller the higher the super Gaussian order $n$. Less intensity variance implies less variation of the differential AC Stark shift and consequently less spread in Rabi oscillation amplitudes, leading to the observed reduction in $F=4$ population spread (see Fig.~\ref{fig.Raman_superGaussian}), and frequencies, resulting in longer decay times. The results for this scenario are better than those for a 5-$\mu$K atom sample in a well-aligned Gaussian beam in all aspects. Thus, the use of super Gaussian addressing lasers can significantly alleviate the  effects of radial misalignment on the Rabi oscillations of an atomic qubit.

Thus far, we have ignored the effects of misalignment on crosstalk. At $1~\mu$m of radial misalignment, the crosstalk intensity at the location of a neighboring site is higher than that of an aligned Gaussian beam. For direct comparison, we also carried out the Rabi oscillation simulations for 150~nm radial misalignment, the amount for which all super Gaussian orders have the same crosstalk intensity at a neighboring site as an aligned Gaussian beam, and as each other, due to our choice of beam waists (see Table~\ref{tab:parameters}). We found the following decay times: 25(3)~ms for $n=4$, 69(12)~ms for $n=6$, and 70(14)~ms for $n=8$. Thus, with the same crosstalk intensity, the decay time for the $n=6$ super Gaussian addressing beam is over 60 times longer than that of a well-aligned Gaussian beam (see Table~\ref{tab:RabiTdepGSG}).

Finally, we explored how sensitive the Rabi oscillations are to axial beam misalignment. Both the atom confinement in the Gaussian beam array trap and the Raman beam alignment precision may be  worse by a factor of approximately ten in the axial direction, so we calculated the Rabi oscillations for an atom temperature of 20~$\mu$K for axial beam misalignments of 2.5~$\mu$m and 5~$\mu$m for Gaussian beams, and also examined super Gaussian beams at 5~$\mu$m axial misalignment. The results are listed in Table~\ref{tab:RabimisalignaxialGSG}.

\begin{table}%[width=86mm]%[H] add [H] placement to break table across pages
%\begin{ruledtabular}
\begin{tabular}{c  c  c  c  c  c  c  c  c}
\hline
\hline
$n$ & $\Delta z$ & $\frac{\Delta_{acR}}{2\pi}$ & $\frac{\Omega}{2\pi}$ & $\frac{\Omega'}{2\pi}$ & $t_a$ & $t_a$ & $\pi$ pop. & 3$\pi$ pop.  \\
 & ($\mu$m) & (kHz) & (kHz) & (kHz) & (ms) & ($\pi$ pulse time $\times$100) & ($\%$) & ($\%$) \\
\hline
2 & 2.5 & -13.06(2) & 22.64(4) & 22.57(4) & 1.13(14) & 0.51(6) & 99.79(2) & 98.98(14) \\
2 & 5 & -12.91(4) & 22.39(7) & 22.31(6) & 1.5(3) & 0.69(14) & 99.67(5) & 98.9(2) \\
4 & 5 & -14.06(3) & 24.38(6) & 24.34(2) & 1.8(2) & 0.87(9) & 99.62(5) & 98.5(2) \\
6 & 5 & -13.479(12) & 23.37(2) & 23.375(11) & 19(5) & 9(2) & 99.87(2) & 99.82(8) \\
8 & 5 & -13.53(2) & 23.45(4) & 23.46(2) & 1.6(3) & 0.76(14) & 99.68(7) & 98.7(3) \\
\hline
\hline
\end{tabular}
\setlength{\tabcolsep}{36pt}
%\end{ruledtabular}
\caption{Rabi oscillation results for atoms with a temperature of 20~$\mu$K in Gaussian ($n=2$) and super Gaussian ($n=4,6,8$) beams with axial misalignment $\Delta z=2.5~\mu$m ($n=2$) and 5~$\mu$m ($n=2, 4, 6,$ and $8$) between the trap site and the focus of the 459-nm wavelength addressing laser beams. The parameters listed are the differential AC Stark shift $\Delta_{acR}$ due to the Raman lasers, the effective on-resonance ($\Omega$) and off-resonance ($\Omega'$) Rabi frequencies, the 1/e Rabi oscillation amplitude decay time $t_a$, and the $F=4$ population for $\pi$ and 3$\pi$ Raman pulse times, respectively. For an atom at the center of the trap and the addressing laser beams $\Delta_{acR}=-2\pi\times13.49$~kHz and $\Omega=2\pi\times23.40$~kHz. The average contribution to the differential AC Stark shift due to the trap laser is $\Delta_{acT}=2\pi\times0.809(6)$~kHz.}
\label{tab:RabimisalignaxialGSG}
\end{table}

For a Gaussian beam, we find that as the axial misalignment is increased to 2.5~$\mu$m and then 5~$\mu$m, there is a small drop in differential Stark shift and Rabi frequencies, which can be explained by the small drop in intensity of the Gaussian beam in this axial range (see Fig.~\ref{fig:GSGonaxis}). The increase in intensity variance (see Fig.~\ref{fig:variancevsz}) is so gradual that its effects on the decay time and the $F=4$ populations at $\pi$ and 3$\pi$ times are small to insignificant. We conclude that Gaussian addressing beams are not very sensitive to axial misalignment.

For super Gaussian Raman beams, the results follow the same trends as those of the perfectly aligned case for the same reasons. As shown in Fig.~\ref{fig:GSGonaxis}, the $n=4$ super Gaussian has a higher intensity at $z=5~\mu$m than at the focal plane, leading to a larger differential Stark shift and increased Rabi frequencies, both compared to the Gaussian beam and the aligned case. The $n=6$ super Gaussian beam has a flat axial intensity profile, resulting in differential Stark shifts and Rabi frequencies closest to those for an atom at the center of the trap for aligned Raman laser beams. The $n=8$ super Gaussian axial intensity profile is oscillating about the focal plane intensity, leading to results very close to those for an atom at the center for aligned addressing beams. The decay time of the Rabi oscillation amplitude and the $F=4$ populations for orders $n=2,4,$ and $8$ match within their uncertainties, while those for $n=6$ are significantly larger. The decay time for $n=6$ is approximately ten times longer than those for the other orders, stemming from its lower intensity variance as shown in Fig.~\ref{fig:variancevsz} and thus narrower range of Rabi frequencies. 

Overall, these results indicate, that for the cases explored here, the $n=6$ super Gaussian beam would be the ideal choice for Raman addressing beams, least sensitive against radial and axial misalignment, providing the most even intensity profile for atoms moving around in the Gaussian beam array trap, resulting in Rabi oscillations with long decay times and little spread in $F=4$ population.

We also repeated the simulations for Raman laser beams with a 100~GHz detuning from the $6^2S_{1/2},F=3,4\rightarrow7^2P_{1/2},F=4$ transitions, respectively, and five times the laser power. This keeps the on-resonance Rabi frequency approximately the same while reducing the differential AC Stark shift due to the Raman lasers by about a factor of six. The differential AC Stark shifts, Rabi frequencies, Rabi oscillation amplitude decay times, $F=4$ populations at $\pi$ and 3$\pi$ times, and spread in $F=4$ population follow the same trends with temperature and axial misalignment as the 20~GHz results. However, the effects of radial misalignment on the decay time and the spread of the $F=4$ population were significantly reduced. This confirms that the spread in Rabi oscillation amplitudes (leading to population spread) and frequencies (resulting in Rabi oscillation decay) is caused by the spread in the differential AC Stark shift, since that is lower by a factor of six in the 100~GHz case. This means that an increase in laser power and corresponding increase in laser detuning is also a way to improve the consistency of the resulting quantum operations. However, laser power is limited, so increasing the detuning may not be possible in a given experimental setup.

Thus far, we have assumed that the Raman laser beams are free from any laser power noise. We repeated each simulation for 2$\%$ laser power noise on the Raman laser beams. For each atom in the simulation, a laser power number in the $\pm1\%$ range is randomly generated and stays constant through the whole simulation. We thus are assuming that the laser noise varies slowly compared to the timescales investigated here (0.15~ms). We found that laser power noise has little to no effect on the differential AC Stark shift due to the Raman lasers, the Rabi frequencies, and the $F=4$ populations at $\pi$ and 3$\pi$ pulse times. However, in some cases the decay times were reduced by factors of two or three with stronger effects for cases with larger decay times. Since the most strongly affected cases were those with large decay times, this reduction does not have a significant effect on the overall performance of quantum operations on the $\pi$ to 3$\pi$ pulse time scales. Thus, the effects of noise are negligible compared to the effects of atom temperature and beam misalignment.

Overall, we found that while atom temperature and laser noise impact the Rabi oscillations, the most significant factor influencing the effective off-resonant Rabi frequency, Rabi oscillation amplitude decay time, and $F=4$ population spread is the Raman beam alignment. Super Gaussian laser beams have a smaller intensity variance, which significantly reduces the sensitivity of the quantum operation performed to alignment. Therefore, the use of super Gaussian Raman addressing beams may help reduce downtime for repeated beam alignment and improve the consistency of quantum operations. Specifically, the flat radial and axial intensity profile of the $n=6$ super Gaussian beam make it ideal as an addressing laser beam for quantum operations on atomic qubits.

\section{Conclusions}
\label{sec:conclusions}

We have presented a detailed parametric study of the effects of finite temperature, radial and axial beam misalignment, and laser noise on the fidelity of stimulated Raman single qubit gates. Our results show that a high order super-Gaussian beam provides more uniform intensity, and greatly increased coherence of Rabi oscillations, as presented in Tables \ref{tab:RabiTdepGSG}-\ref{tab:RabimisalignaxialGSG}, together with  less crosstalk for optically trapped atomic qubits. 

The optimal choice of the super Gaussian index $n$ will depend on details of the experimental environment. While the super Gaussian beams provide lower intensity variance and less crosstalk for small axial or radial misalignment they perform worse than a Gaussian beam for large misalignment. For radial misalignment of not more than  $\sim 150 ~\rm nm$ and axial misalignment of not more than $\sim 2~\mu\rm m$ we find a $n=6$ super Gaussian to be near optimal. These misalignment values are realistic estimates based on recent experiments\cite{Xia2015,Maller2015a}. For larger misalignment  values  a lower order should be chosen as can be seen from 
Fig. \ref{fig:GSGxtalkrandz}. On the other hand, if it is possible to ensure very small beam misalignment then the lowest possible intensity variance, and best possible Rabi oscillation fidelity,  is obtained for $n=6.5$ as is shown in 
Table \ref{tab:variancealignednear6}. Variations of $n$ by $\pm0.5$ about this value have only a small effect on the intensity variance. 

It is apparent from our results that the super Gaussian beam has only a minor effect on the amplitude achieved in a $\pi$ or $3\pi$ pulse, but has a large influence, by up to a factor of 60, on the decay time of the Rabi oscillations. Very high fidelity gates as needed for scalable quantum computation  will likely be based on composite pulse techniques\cite{Mount2015} to minimize sensitivity to imperfect control of the pulse area. Since composite pulses imply larger total pulse areas the increase of the decay time afforded by the super Gaussian pulse will be particularly  useful in conjunction with  composite pulses.

We have studied the super Gaussian beam profile because it has a compact analytical form and is readily visualized. This leaves open the question of what the global optimal beam shape might be. The choice of an optimal shape will strongly depend on the experimental conditions. If there is no experimental beam misalignment and the atoms are very cold and well localized then a high order beam profile with excellent uniformity is to be preferred. Allowing for misalignment and atomic motion a lower order profile with smoothly varying intensity gives better performance. The specific choice can not be predicted in general and will depend on actual parameters.

In order to reap full advantage of using a specific beam profile it is of course necessary to generate such a profile experimentally. There are several methods available for beam shaping of top hat, or similar profiles,  including aspherical lenses\cite{Hoffnagle2000}, diffractive optical 
elements\cite{Han1983,Reetz-Lamour2008a}, and computer controlled spatial light modulators (SLMs)\cite{Pasienski2008}. Although arbitrary beam shapes with desired amplitude and phase profiles can in principle be produced using a SLM in holographic mode there remains the experimental challenge of compensating for imperfections due to optical elements and vacuum windows. A notable advance was reported recently in \cite{Nogrette2014} where a SLM together with a Shack-Hartmann wave-front sensor was used to correct for imperfections in the optical train and obtain a uniform array of intensity spots with a standard deviation of only 1.4\%. Based on these techniques it should be possible to create essentially any desired beam profile with accuracy 
at the 1\% level. Such accuracy together with the weak sensitivity of the atom averaged intensity variance with super Gaussian index presented in Table \ref{tab:variancealignednear6}, suggests that performance comparable to that calculated here will be possible in real experiments.

\begin{acknowledgement}
MS and MJP were supported by the IARPA MQCO program through ARO contract W911NF-10-1-0347.
\end{acknowledgement}

\appendix*

\section{On-resonance Rabi frequency and differential AC Stark shift for two-photon Raman transitions via $7^2P_{1/2}$ in $^{133}$Cs including hyperfine splitting}\label{sec:CsRabifandStark}

The Rabi oscillations investigated in this work are driven via a Raman process from the $F=3, m_F=0$ hyperfine ground state of the $6^2S_{1/2}$ manifold in $^{133}$Cs to its $F=4,m_F=0$ hyperfine ground state via the $7^2P_{1/2}$ manifold using two laser beams. To treat this kind of Rabi oscillation, we repeat the steps from Sec.~\ref{sec:Rabioscillations} for a $\Lambda$ type three-level system with two lasers tuned to the two transitions of the Raman process. For detunings large enough so that the excited state population is small, we can adiabatically eliminate the $7^2P_{1/2}$ state, resulting in an effective two-level Rabi oscillation with an on-resonance Rabi frequency of
\begin{equation}
\Omega=\frac{\Omega_1 \Omega_2^*}{2 \Delta_R},
\label{eq:2photonRabif}
\end{equation}
where $\Omega_{1,2}$ are the single photon on-resonance Rabi frequencies for the $6^2S_{1/2}, F=3, m_F=0 \rightarrow 7^2P_{1/2}$ and $7^2P_{1/2} \rightarrow 6^2S_{1/2}, F=4, m_F=0$ transitions, respectively. $\Delta_R$ is the detuning of the first Raman laser beam from the $6^2S_{1/2}, F=3 \rightarrow 7^2P_{1/2}$ (fine structure level) transition, and we have assumed that the detuning of the second Raman laser from the $6^2S_{1/2}, F=4 \rightarrow 7^2P_{1/2}$ transition is the same. Equation (\ref{eq:2photonRabif}) is valid for two-photon resonance or when the departure from two-photon resonance is small compared to $\Delta_R$.

Taking into account the hyperfine splitting of the $7^2P_{1/2}$ level, we have to sum over all possible intermediate states, resulting in
\begin{equation}
\Omega=\sum_{F'}\frac{\Omega_{1,F_1 F'} \Omega_{2,F' F_2}^*}{2 \Delta_{R,F'}},
\nonumber%\label{eq:2photonRabifhyperfine}
\end{equation}
where $F', F_{1,2}$ are the total angular momentum quantum numbers of the intermediate, initial, and final states, respectively, $\Omega_{1,F_1 F'}$ and $\Omega_{2,F' F_2}$ are the single photon on-resonance Rabi frequencies for the $F_1 \rightarrow F'$ and $F' \rightarrow F_2$ transitions, respectively, and $\Delta_{R,F'}$ is the detuning of the first Raman laser beam from the $F_1 \rightarrow F'$ transition.

In $^{133}$Cs, we have $F_1=3$, $F_2=4$, and $F'=3,4$. We thus find for the two-photon on-resonance Rabi frequency
\begin{equation}
\Omega=\frac{\Omega_{1,33'} \Omega_{2,3'4}^*}{2 \Delta_{R,3'}}+\frac{\Omega_{1,34'} \Omega_{2,4'4}^*}{2 \Delta_{R,4'}},
\nonumber%\label{eq:2photonRabifhyperfineCs}
\end{equation}
where we used primes to indicate quantum numbers pertaining to the excited states. 
The detunings from the $7^2P_{1/2}$ hyperfine states are $\Delta_{R,3'}=\Delta_R-\Delta_{F'3}$ and $\Delta_{R,4'}=\Delta_R-\Delta_{F'4}$. Here, $\Delta_{F'3}=-2\pi\times212.3~\text{MHz}$ and $\Delta_{F'4}=2\pi\times165.1~\text{MHz}$ are the hyperfine shifts from the $7^2P_{1/2}$ fine structure level to the $F'=3,4$ hyperfine states, respectively.

The one-photon Rabi frequencies are $\Omega_{i,F_i F_f}=\Omega_{i,0} \tilde\Omega_{F_i F_f}$
with $\Omega_{i,0}= {\mathcal E}_i e \langle 7^2P_{1/2}||r||6^2S_{1/2}\rangle/\hbar$, where ${\mathcal E}_i$ is the electric field amplitude of Raman laser $i=1,2$, $e\langle 7^2P_{1/2}||r||6^2S_{1/2}\rangle$ is the reduced dipole matrix element for the $6^2S_{1/2} \rightarrow 7^2P_{1/2}$ transition, $e$ is the elementary charge, and
\begin{equation}
\tilde{\Omega}_{F_1 F'}=c_{J_1,I,F_1}^{J',I,F'}C_{F_1,m_{F1},1,q_1}^{F',m_{F1}+q_1}
\nonumber%\label{eq:Omegatildeabs}
\end{equation}
for a Raman absorption and
\begin{equation}
\tilde{\Omega}_{F' F_2}=c_{J',I,F'}^{J_1,I,F_2}C_{F',m_{F1}+q_1,1,-q_2}^{F_2,m_{F1}+q_1-q_2}
\nonumber%\label{eq:Omegatildeem}
\end{equation}
for a stimulated Raman emission. Here, $C_{F_1,m_{F1},1,q_1}^{F',m_{F1}+q_1}$ and $C_{F',m_{F1}+q_1,1,-q_2}^{F_2,m_{F1}+q_1-q_2}$ are Clebsch-Gordan coefficients, and 
\begin{equation}
c_{J_1,I,F_i}^{J',I,F_j}=(-1)^{1+I+F_i+J'}\sqrt{2F_i+1}\left\{\begin{array}{ccc} J_1 & I & F_i \\
F_j & 1 & J' \end{array}\right\}.
\nonumber%\label{eq:ccoefficients}
\end{equation}

For our specific transitions in $^{133}$Cs, $6^2S_{1/2}, F=3, m_F=0 \rightarrow 7^2P_{1/2}, F=3,4, m_F=1$ and $7^2P_{1/2}, F=3,4, m_F=1 \rightarrow 6^2S_{1/2}, F=4, m_F=0$, the relevant quantum numbers are the initial and final total electron angular momentum quantum numbers $J_1=J'=1/2$, the nuclear spin quantum number $I=7/2$, the initial magnetic quantum number $m_{F1}=0$, and we use circularly polarized Raman laser beams such that the z-components of the angular momentum of the absorbed and emitted photons are $q_1=q_2=1$. With these, we find
\begin{eqnarray}
\tilde\Omega_{33'}&=&c_{1/2,7/2,3}^{1/2,7/2,3}C_{3,0,1,1}^{3,1}=1/4,\nonumber\\
\tilde\Omega_{3'4}&=&c_{1/2,7/2,3}^{1/2,7/2,4}C_{3,1,1,-1}^{4,0}=1/4,\nonumber\\
\tilde\Omega_{34'}&=&c_{1/2,7/2,3}^{1/2,7/2,4}C_{3,0,1,1}^{4,1}=\sqrt{5/48},\nonumber\\
\tilde\Omega_{4'4}&=&c_{1/2,7/2,4}^{1/2,7/2,4}C_{4,1,1,-1}^{4,0}=\sqrt{5/48}.\nonumber
\end{eqnarray}

The reduced dipole matrix element for the $6^2S_{1/2}\rightarrow 7^2P_{1/2}$ transition in $^{133}$Cs is $e\langle 7^2P_{1/2}||r||6^2S_{1/2}\rangle=0.276 ea_0$ \cite{Iskrenova-Tchoukova2007}, where $a_0$ is the Bohr radius.

Altogether, we find the two-photon on-resonance Rabi frequency to be
\begin{eqnarray}
\Omega
&=&\frac{\Omega_{1,0}\Omega_{2,0}^*}{32}\left(\frac{1}{ \Delta_R-\Delta_{F'3}}+\frac{5/3}{\Delta_R-\Delta_{F'4}}\right).
\nonumber%\label{eq:2photonRabiffinal}
\end{eqnarray}

To find the total Stark shift of each of the hyperfine ground states, we need to add the Stark shifts due to the first and second Raman laser beams, so
\begin{equation}
\Delta_{ac,F}=\Delta_{ac,F, R1}+\Delta_{ac,F, R2}
\label{eq:StarkF}
\end{equation} 

For each hyperfine ground state, we must sum over the contributions due to each of the hyperfine states of the $7^2P_{1/2}$ manifold, resulting in
\begin{equation}
\Delta_{ac,F, R1/R2}=\sum_{F'}\frac{|\Omega_{F F'}|^2}{4\Delta_{R,F'}}.
\nonumber%\label{eq:StarkRamansum}
\end{equation}

Thus, the AC Stark shift of the $F=3$ hyperfine ground state due to the first Raman laser is
\begin{eqnarray}
\Delta_{\rm ac,3, R1}&=&\frac{|\Omega_{1,33'}|^2}{4\Delta_{R,3'}}
+\frac{|\Omega_{1,34'}|^2}{4\Delta_{R,4'}}\nonumber\\
&=&\frac{|\Omega_{1,0}|^2}{64}\left(\frac{1}{\Delta_R-\Delta_{F'3}}+
\frac{5/3}{\Delta_R-\Delta_{F'4}}\right).
\nonumber%\label{eq:StarkF3Raman1}
\end{eqnarray} 

Similarly, the contribution to the AC Stark shift due to the second Raman laser is
\begin{eqnarray}
\Delta_{\rm ac,3,R2}&=&\frac{|\Omega_{2,33'}|^2}{4\left(\Delta_{R,3'}-\Delta_{hf}\right)}+
\frac{|\Omega_{2,34'}|^2}{4\left(\Delta_{R,4'}-\Delta_{hf}\right)}\nonumber\\
&=&\frac{|\Omega_{2,0}|^2}{64}\left(\frac{1}{\Delta_R-\Delta_{F'3}-\Delta_{hf}}+
\frac{5/3}{\Delta_R-\Delta_{F'4}-\Delta_{hf}}\right),
\nonumber%\label{eq:StarkF3Raman2}
\end{eqnarray} 
where $\Delta_{hf}=2\pi\times9.192631770$~GHz is the ground state hyperfine splitting of $^{133}$Cs. 

The contribution of the first Raman laser to the AC Stark shift of the $F=4$ hyperfine ground state is
\begin{eqnarray}
\Delta_{\rm ac,4,R1}&=&\frac{|\Omega_{1,43'}|^2}{4\left(\Delta_{R,3'}+\Delta_{hf}\right)}+
\frac{|\Omega_{1,44'}|^2}{4\left(\Delta_{R,4'}+\Delta_{hf}\right)}\nonumber\\
&=&\frac{|\Omega_{1,0}|^2}{64}\left(\frac{1}{\Delta_R-\Delta_{F'3}+\Delta_{hf}}+
\frac{5/3}{\Delta_R-\Delta_{F'4}+\Delta_{hf}}\right).
\nonumber%\label{eq:StarkF4Raman1}
\end{eqnarray}

Here, we have used
\begin{equation}
\tilde\Omega_{43'}=c_{1/2,7/2,4}^{1/2,7/2,3}C_{4,0,1,1}^{3,1}=-1/4.
\nonumber%\label{eq:Omega43prime}
\end{equation}
and
\begin{equation}
\tilde\Omega_{44'}=c_{1/2,7/2,4}^{1/2,7/2,4}C_{4,0,1,1}^{4,1}=-\sqrt{5/48}.
\nonumber%\label{eq:Omega44prime}
\end{equation}

Finally, the contribution of the second Raman laser beam to the AC Stark shift of the $F=4$ ground state is
\begin{eqnarray}
\Delta_{\rm ac,4,R2}&=&\frac{|\Omega_{2,43'}|^2}{4\left(\Delta_{R,3'}\right)}+
\frac{|\Omega_{2,44'}|^2}{4\left(\Delta_{R,4'}\right)}\nonumber\\
&=&\frac{|\Omega_{2,0}|^2}{64}\left(\frac{1}{\Delta_R-\Delta_{F'3}}+
\frac{5/3}{\Delta_R-\Delta_{F'4}}\right).
\nonumber%\label{eq:StarkF4Raman2}
\end{eqnarray} 

The differential Stark shift between the $F=3$ and $F=4$ hyperfine ground states is
\begin{equation}
\Delta_{acR}=\Delta_{ac,4}-\Delta_{ac,3}.
\nonumber%\label{eq:diffStark1}
\end{equation}

In this work, we used two Raman laser beams of identical power, waist, and alignment, so ${\mathcal E}_1={\mathcal E}_2$, and consequently
\begin{eqnarray}
\Delta_{acR}=\frac{|\Omega_{1,0}|^2}{64} \left(\frac{1}{\Delta_R-\Delta_{F'3}+\Delta_{hf}}\right.&+&\left.
\frac{5/3}{\Delta_R-\Delta_{F'4}+\Delta_{hf}}\right.\nonumber\\
-\left.\frac{1}{\Delta_R-\Delta_{F'3}-\Delta_{hf}}\right.&-&\left.
\frac{5/3}{\Delta_R-\Delta_{F'4}-\Delta_{hf}}\right).
\nonumber%\label{eq:diffStarkfinal2}
\end{eqnarray}

%\bibliographystyle{iopart-num}
%\bibliography{supergaussian} 

%%%%%%%%%%%%%%%%%%%%%
%%% bibliography from bibtex created .bbl file 
%%%%%%%%%%%%%%%%%%%%%%%

\providecommand{\newblock}{}

\end{document}